\documentclass[11pt]{article}

\newlength{\vshift}
\newlength{\hshift}
\usepackage{amssymb,amsopn}

\renewcommand{\theequation}{\thesection.\arabic{equation}}
\newcommand{\initiate}{\setcounter{equation}{0}}

\catcode`\@=11
\newdimen\cdsep
\def\cdstrut{\vrule height .6\cdsep width 0pt depth .4\cdsep}
\def\@cdstrut{{\advance\cdsep by 2em\cdstrut}}

\def\arrow#1#2{
  \ifx d#1
    \llap{$\scriptstyle#2$}\left\downarrow\cdstrut\right.\@cdstrut\fi
  \ifx u#1
    \llap{$\scriptstyle#2$}\left\uparrow\cdstrut\right.\@cdstrut\fi
  \ifx r#1
    \mathop{\hbox to \cdsep{\rightarrowfill}}\limits^{#2}\fi
  \ifx l#1
    \mathop{\hbox to \cdsep{\leftarrowfill}}\limits^{#2}\fi
}
\catcode`\@=12

\def\nn{\nonumber }

\def\be{\beta}
\def\a{\alpha}

\def\g{\gamma}

\def\ds{\stackrel{\star}{,}}

\def\t{\tilde}

\def\diff{\textrm{d} }

\def\Str{{\rm STr}}
\def\tr{{\rm Tr}}

\def\SF{\mathbb{F}}

\def\nn{\nonumber}
\def\be{\begin{equation}}             \def\ee{\end{equation}}
\def\ba#1{\begin{array}{#1}}          \def\ea{\end{array}}
\def\bea{\begin{eqnarray} }           \def\eea{\end{eqnarray} }
\def\beann{\begin{eqnarray*} }        \def\eeann{\end{eqnarray*} }
\def\beal{\begin{eqalign}}            \def\eeal{\end{eqalign}}
             
\def\bsubeq{\begin{subequations}}     \def\esubeq{\end{subequations}}
\def\bitem{\begin{itemize}}           \def\eitem{\end{itemize}}
                    
\def\pa{\partial}
\def\a{\alpha}
\def\b{\beta}

\def\e{\epsilon}                % Also, \varepsilon
\def\g{\gamma}

\def\k{\kappa}
\def\l{\lambda}
\def\m{\mu}
\def\n{\nu}
\def\o{\omega}
\def\r{\rho}                    %       \varrho
\def\s{\sigma}                  %       \varsigma
\def\t{\tau}

\def\G{\Gamma}

\def\L{\Lambda}
\def\O{\Omega}

\def\S{\Sigma}

\usepackage{mathrsfs}
\usepackage{mathtools}
\usepackage{amsmath}
\usepackage{xcolor}
\usepackage{tikz}
\usepackage[numbers,sort&compress]{natbib}

\usepackage[utf8]{inputenc}

\makeatletter
\@addtoreset{equation}{enumi}
\makeatother

\usepackage{slashed}
\usepackage{scalerel}
\usepackage{stackengine}
\stackMath
\newcommand\reallywidehat[1]{%
\savestack{\tmpbox}{\stretchto{%
  \scaleto{%
    \scalerel*[\widthof{\ensuremath{#1}}]{\kern-.6pt\bigwedge\kern-.6pt}%
    {\rule[-\textheight/2]{1ex}{\textheight}}%WIDTH-LIMITED BIG WEDGE
  }{\textheight}%
}{0.5ex}}%
\stackon[1pt]{#1}{\tmpbox}%
}
\newcommand\myeq{\stackrel{\text{WI}}{=}}
\newcommand\myeqe{\stackrel{\text{g.f.}}{=}}

\begin{document}
\vspace{1.5cm}
\begin{titlepage}

\begin{center}

{\LARGE{\bf Canonical Deformation of $N=2$ $AdS_{4}$ SUGRA }}

\vspace*{1.3cm}

{{\bf Dragoljub Go\v{c}anin and Voja Radovanovi\'{c}}}

\vspace*{1cm}

University of Belgrade, Faculty of Physics\\
Studentski trg 12, 11000 Beograd, Serbia \\[1em]

\end{center}

\vspace*{2cm}

\date{\today}

\begin{abstract}
It is well known that one can define a consistent theory of extended, $N=2$ anti-de Sitter (AdS) Supergravity (SUGRA) in $D=4$. Besides the standard gravitational part (including negative cosmological constant), this theory involves a single $U(1)$ gauge field and a pair of Majorana vector-spinors that can be combined to form a pair of Dirac vector-spinors (charged spin-$3/2$ gravitini). The action for $N=2$ $AdS_{4}$ SUGRA is invariant under $SO(1,3)\times U(1)$ gauge transformations, and under local complex SUSY. We present a geometric
action that involves two ``inhomogeneous'' parts: an orthosymplectic $OSp(4\vert 2)$ gauge-invariant action of the Yang-Mills type, and a supplementary action invariant under purely bosonic $SO(2,3)\times U(1)\sim Sp(4)\times SO(2)$ sector of $OSp(4\vert 2)$, that needs to be added for consistency. This action reduces to $N=2$ $AdS_{4}$ SUGRA after suitable gauge fixing, for which we use a constrained auxiliary field in the manner of Stelle and West.  
Canonical ($\theta$-constant) deformation is performed by using the Seiberg-Witten approach to noncommutative (NC) gauge field theory with the Moyal $\star$-product. The NC-deformed action is expanded in powers of the deformation parameter $\theta^{\m\n}$, up to the first order. We show that $N=2$ $AdS_{4}$ SUGRA has non-vanishing linear NC correction in the physical gauge, originating from the additional, purely bosonic action. 
For comparison, simple $N=1$ Poinacar\'{e} SUGRA can be obtained in the same manner from an $OSp(4\vert 1)$ gauge-invariant action (without introducing any additional terms). The first non-vanishing NC correction is quadratic in the deformation parameter $\theta^{\m\n}$, and therefore exceedingly difficult to calculate. 
Under Wigner-In\"{o}n\"{u} contraction, $N=2$ AdS superalgebra reduces to $N=2$ Poincar\'{e} superalgebra, and it is not clear whether this relation holds after canonical NC deformation. 
We present the linear NC correction to $N=2$ $AdS_{4}$ SUGRA explicitly, discuss its low-energy limit, and what remains of it after Wigner-In\"{o}n\"{u} contraction.

\end{abstract}
\vspace*{1cm}

{\bf Keywords:} {Moyal $\star$-product, $OSp(4|2)$ supergroup, NC SUGRA}

\vspace*{1cm}
\quad\scriptsize{e-mail:
dgocanin@ipb.ac.rs, rvoja@ipb.ac.rs}
\vfill

\end{titlepage}

\setcounter{page}{1}
\newcommand{\Section}[1]{\setcounter{equation}{0}\section{#1}}
\renewcommand{\theequation}{\arabic{section}.\arabic{equation}}

%%%%%%%%%%%%%%%%%%%%%%%%%%%%%%%%%%%%%%
\section{Introduction}
\hspace{0.4cm} 
%%%%%%%%%%%%%%%%%%%%%%%%%%%%%%%%%%%%%%
In our quest for the theory of ``Quantum Gravity'', we must be prepared to go beyond some deeply rooted assumptions on which we are accustomed, in particular, at very short distances (very high energies) we might have to abandon the notion of a continuous space-time and the associated mathematical concept of a smooth manifold that describes it. One distinguished approach to the problem is \textit{Noncommutative (NC) Field Theory} - a theory of relativistic fields on \textit{noncommutative space-time}, based on the method of deformation quantization by NC $\star$-product \cite{Flato, Maxim, NCbook}. One speaks of a deformation of an object/structure whenever there is a family of similar objects/structures whose ``distortion'' from the original, ``undeformed'' one, can be somehow parametrized. In physics, this so-called \textit{deformation parameter} appears as some fundamental constant of nature that measures the deviation from the classical (i.e. undeformed) theory. This way of ``quantizing'' space-time is essentially different from the standard QFT quantization procedure for matter fields.  Different space-time dimensions (the usual $3+1$) are regarded as being mutually ``incompatible'', in a sense that there exist a lower bound for the product of uncertainties $\Delta x^{\m}\Delta x^{\n}$ for a pair of two different coordinates. To capture this ``pointlessness'' of space-time, one introduces an abstract algebra of NC coordinates as a deformation of the classical structure. These NC coordinates, denoted by $\hat{x}^{\m}$, satisfy some non-trivial commutation relations, and so, it is no longer the case that $\hat{x}^{\m}\hat{x}^{\n}=\hat{x}^{\n}\hat{x}^{\m}$. Abandoning this basic property of space-time leads to various new physical effects that were not present in theories based on classical space-time. The simplest case of noncommutativity is the so-called \textit{canonical (or $\theta$-constant) noncommutativity},   
\be [\hat{x}^\m, \hat{x}^\n]=i\theta^{\m\n}\sim\L_{NC}^{2}\ ,\label{can-kom-rel}\ee
where $\theta^{\m\n}$ are components of a \textit{constant} antisymmetric matrix,
%\footnote{More precisely, $[\hat{x}^\m, \hat{x}^\n] =i\kbar\theta^{\m\n}$, with deformation parameter $\kbar\sim\L_{NC}^{2}$ and arbitrary constant antisymmetric matrix elements $\theta^{\m\n}$. Conventionally, $\kbar$ is absorbed in the matrix elements $\theta^{\m\n}$ and these we call deformation parameters.}
and $\L_{NC}$ is the length scale at which NC effects become relevant. Deformation parameter is a fundamental constant, like the Planck length or the speed of light.
%%%%%%%%%%%%%%%%%%%%%%%%%%%%%%%%%%%%%%
%%%%%%%%%%%%%%%%%%%%%%%%

Instead of deforming abstract algebra of coordinates one can take an alternative, but equivalent approach in which noncommutativity appears in the form of NC products of functions (fields) of \textit{commutative} variables (coordinates). These products are called \textit{star products} ($\star$-products). In particular, to establish \textit{canonical noncommutativity}, we use the Moyal $\star$-product,  
\begin{equation}
\label{moyal}  (\hat{f}\star\hat{g})(x) =
      e^{\frac{i}{2}\,\theta^{\m\n}\frac{\pa}{\pa x^\m}\frac{\pa}{ \pa
      y^\n}} f (x) g (y)|_{y\to x}\ .
\end{equation}
The leading term in the expansion of the exponential is the ordinary commutative product of functions, and the higher order terms represent NC (non-classical) corrections. 
%%%%%%%%%%%%%%%%%%%%%%%%%%%%%%%%%%%%%%%%%%%%%%%%%%%%%%%%%%%%%%%

Up to date, we still lack direct physical evidence of Supersymmetry (SUSY), at least in its simplest form. Nevertheless, its beneficial influence on high-energy physics (improved renormalizability in QFT and a natural resolution of the hierarchy problem), along with its mathematical consistency and unification power (especially unification of gravity and the Standard Model within Supergravity (SUGRA), and ultimate unification scheme such as Superstring theory), motivate us to seriously consider SUSY as an integral part of our description of nature. Since the pioneering work of Freedman, van Nieuwenhuizen and Ferrara \cite{SUGRA-FNF1, SUGRA-FNF2}, and Deser and Zumino \cite{SUGRA-DZ}, the theory of supergravity has become a well-developed field of research. SUGRA provides a natural unification of gravity with other fields by imposing gauge principle on SUSY, the associated gauge field being the spin-$3/2$ gravitino field described by a Majorana vector-spinor. It was demonstrated in \citep{N=2_AdS_SUGRA1, N=2_AdS_SUGRA2} that one can have a consistent theory of extended $N=2$ $AdS_{4}$ SUGRA with complex ($U(1)$-charged) gravitino field. In this paper, we propose a geometric way of obtaining $N=2$ $AdS_{4}$ SUGRA action and perform its NC deformation. The obtained NC correction can be regarded as a low-energy signature of the underlying theory of quantum gravity. We calculate the correction explicitly and discuss some of its properties.  

The results of this paper amount to a supersymmetric extension of the theory of NC gravity whose various aspects have been treated extensively in the literature \citep{TwistApproach1, TwistApproach2, Chaichian, SWmapApproach1, SWmapApproach2, SWmapApproach3, Others1, Others2, Others3, Others4, Others5, Others6, Others7, symplectic1, symplectic2}. In particular, an approach based on NC-deformed AdS gauge group $SO(2,3)$ is developed in \cite{MiAdSGrav, MDVR-14, UsLetter, Us-16}, building on the results of MacDowell and Mansouri \cite{MacDowell-Mansouri}, Towsend \cite{Towsend}, Stelle and West \cite{stelle-west}, Mukhanov and Chamseddine \cite{Mukhanov1, Mukhanov2} and Wilczek \cite{Wilczek}. One starts with a classical (undeformed) action invariant under $SO(2,3)$ gauge transformations. To relate AdS gauge theory with GR, original $SO(2,3)$ gauge symmetry has to be broken to $SO(1,3)$, by gauge fixing. For that matter, a constrained auxiliary field is introduced, as in \cite{stelle-west}, to define the physical gauge. Spin-connection and vierbein are treated on equal footing, as components of the general $SO(2,3)$ gauge field.
The $SO(2,3)$ gauge-invariant action is deformed by introducing Moyal $\star$-product, and expanded in powers of $\theta^{\m\n}$ via Seiberg-Witten (SW) map \cite{SWMap1, SWMap2, SWMap3, SWMap4}. After symmetry breaking, one obtains NC corrections to classical gravity, invariant under $SO(1,3)$ gauge transformations. The first order NC correction vanishes as confirmed by \cite{PLM-13}. The second-order NC correction to GR is found explicitly, and deformed equations of motion are analyzed. It is argued that apparent breaking of diffeomorphism invariance stems from the fact that, by introducing the \emph{canonical} anti-commutation relations between space-time coordinates (\ref{can-kom-rel}), we implicitly set ourselves in a preferred coordinate system - the Fermi inertial frame along a geodesic \cite{Fermi1, Fermi2, Fermi3}.   

  Similarly, one can establish NC SUGRA by gauging an appropriate supergroup \cite{Fre, Ortin, Freedman, Cham1, Cham2, Cham3, Vasiliev, NoGravitini1, NoGravitini2} and subsequently performing canonical deformation. Since pure gravity can be obtained by gauging AdS group $SO(2,3)$, orthosymplectic supergroup $OSp(4\vert 1)$ appears as a natural choice for pure $N=1$ Poincar\'{e} SUGRA. Bosonic sector of $\mathfrak{osp}(4\vert 1)$ superalgebra - symplectic algebra $\mathfrak{sp}(4)$ - is isomorphic to AdS algebra $\mathfrak{so}(2,3)$ that reduces to Poincar\'{e} algebra under Wigner-In{\"o}n{\"u} contraction \cite{Freund-Kaplansky}. The subject of NC SUGRA has been treated in \cite{NCSUGRA1, NCSUGRA2}. Classical action for $OSp(4\vert 1)$ SUGRA presented in \cite{NCSUGRA2} is manifestly invariant under $OSp(4\vert 1)$ gauge transformations, and we will use it as a motivation. However, to obtain explicit NC deformation of this action is exceedingly difficult, because the first non-vanishing NC correction is quadratic in $\theta^{\m\n}$. Taking a lesson from \cite{VG, U(1), Paolo-fer} that inclusion of Dirac spinors coupled to $U(1)$ gauge field produces (much simpler) linear NC correction, we will make a transition to $OSp(4\vert 2)$ SUGRA that involves a pair of Majorana spinors that can be mixed into a pair of charged spin-$3/2$ gravitini coupled to $U(1)$ gauge field. We present a geometric action that consists of two ``inhomogeneous'' parts: an $OSp(4|2)$ gauge-invariant action of the Yang-Mills type, and a supplementary action, invariant under purely bosonic $SO(2,3)\times U(1)$ sector of $OSp(4\vert 2)$, that has to be included in order to obtain complete $N=2$ $AdS_{4}$ SUGRA at the classical level; this additional bosonic term produces a non-trivial linear NC correction to $N=2$ $AdS_{4}$ SUGRA, after deformation. 

In Section $2$, we introduce undeformed geometric action for $OSp(4\vert 2)$ SUGRA and make comparison with the similar action for $OSp(4|1)$ SUGRA. In Section $3$, we perform NC deformation by using the Seiberg-Witten approach, and study the first order NC correction to $N=2$ $AdS_{4}$ SUGRA. Section $4$ contains discussion and proposals for further investigation. Appendices A and B contain supplementary material.

%%%%%%%%%%%%%%%%%%%%%%%%%%%%%%%%%%%%%%%%%%%%%%%%%%%%%%

\newpage
\initiate 
%%%%%%%%%%%%%%%%%%%%%%%%%%%%%%%%%%%%%%%%%%%%%%%%%%%%%%%
\section{Classical Orthosymplectic SUGRA}
%%%%%%%%%%%%%%%%%%%%%%%%%%%%%%%%%%%%%%%%%%%%%%%%%%%%%%%
  We consider two \emph{classical} (i.e. undeformed) SUGRA models based on orthosymplectic $OSp(4\vert N)$ gauge group: the simple $N=1$ $AdS_{4}$ SUGRA, describing pure supergravity with the negative cosmological constant, and the extended $N=2$ $AdS_{4}$ SUGRA that involves also a pair of charged gravitini fields coupled to $U(1)$ gauge field. We focus our attentions on the latter ($N=2$), since the former ($N=1$) has been treated extensively in \cite{NCSUGRA2}, including its NC deformation, and we discuss it just for comparison. Some significant differences of the two models in question are manifested already at the level of their classical actions, and this reflects drastically on the structure of their NC corrections after deformation.  

%%%%%%%%%%%%%%%%%%%%%%%%%%%%%%%%%%%%%%%%%%%
\subsection{Classical $OSp(4\vert 2)$ SUGRA} 
%%%%%%%%%%%%%%%%%%%%%%%%%%%%%%%%%%%%%%%%%%%
  Orthosymplectic group $OSp(4\vert 2)$ has $19$ generators, and they are of two kinds - bosonic and fermionic. Ten bosonic generators $\hat{M}_{AB}=-\hat{M}_{BA}$ ($A,B=0,1,2,3,5$) span AdS Lie algebra $\mathfrak{so}(2,3)$ (symmetry algebra of $AdS_{4}$), 
\be
[\hat{M}_{AB},\hat{M}_{CD}]=i(\eta_{AD}\hat{M}_{BC}+\eta_{BC}\hat{M}_{AD}-\eta_{AC}\hat{M}_{BD}-\eta_{BD}\hat{M}_{AC
})\ ,
\label{AdSalgebra}
\ee
where $\eta_{AB}$ is flat $5$D metric with signature $(+,-,-,-,+)$. By splitting this set of generators into six $\hat{M}_{ab}$ AdS rotation generators ($a,b=0,1,2,3$) and four AdS translation generators $\hat{M}_{a5}$, we can recast $\mathfrak{so}(2,3)$ algebra in a more explicit form:
\begin{align}
[\hat{M}_{a5}, \hat{M}_{b5}]&=-i\hat{M}_{ab}\ ,\nn\\
[\hat{M}_{ab}, \hat{M}_{c5}]&=i(\eta_{bc}\hat{M}_{a5}-\eta_{ac}\hat{M}_{b5})\ ,\nn\\
[\hat{M}_{ab}, \hat{M}
_{cd}]&=i(\eta_{ad}\hat{M}_{bc}+\eta_{bc}\hat{M}_{ad}-\eta_{ac}\hat{M}_{bd}-\eta_{bd}\hat{M}_{ac
})\ . \label{AdS-split}
\end{align}
If we introduce a new set of generators $(\hat{\mathcal{M}}_{ab}, \hat{\mathcal{P}}_{a})$ defined by $\hat{\mathcal{M}}_{ab}:=\hat{M}_{ab}$ and $\hat{\mathcal{P}}_{a}:=l^{-1}\hat{M}_{a5}=\a \hat{M}_{a5}$, where $l$ is a length scale related to AdS radius and $\a=l^{-1}$ (we will use both parameters in the following formulae), the algebra (\ref{AdS-split}) transforms into:
\begin{align}
[\hat{\mathcal{P}}_{a}, \hat{\mathcal{P}}_{b}]&=-i\a^{2}\hat{\mathcal{M}}_{ab}\ , \nn\\
[\hat{\mathcal{M}}_{ab}, \hat{\mathcal{P}}_{c}]&=i(\eta_{bc}\hat{\mathcal{P}}_{a}-\eta_{ac}\hat{\mathcal{P}}_{b})\ ,\nn\\
[\hat{\mathcal{M}}_{ab}, \hat{{\mathcal{M}}}_{cd}]&=i(\eta_{ad}\hat{\mathcal{M}}_{bc}+\eta_{bc}\hat{\mathcal{M}}_{ad}-\eta_{ac}\hat{\mathcal{M}}_{bd}-\eta_{bd}\hat{\mathcal{M}}_{ac
}) \ . \label{AdS-PP}
\end{align}
In the limit $\a\rightarrow 0$ (or $l\rightarrow\infty$), AdS algebra reduces to Poincar\'{e} algebra, in particular, we obtain 
$[\hat{\mathcal{P}}_{a}, \hat{\mathcal{P}}_{b}]=0$ with all other commutators left unchanged. This is a famous example of the Wigner-In{\"o}n{\"u} (WI) contraction, the contraction parameter being $\a$ (or $l$). This Lie-algebra contraction (or deformation) can be extended to AdS superalgebra, and we will be interested, later on, in its effect on the NC correction of $N=2$ $AdS_{4}$ SUGRA. 
%%%%%%%%%%%%%%%%%%%%%%%%%%%%%%%%%%%%%%%%%%%%%%%%%%%%%%%%%%

  A representation of the AdS sector of $\mathfrak{osp}(4\vert 1)$ superalgebra can be obtained by using $5$D gamma matrices $\G_A$ satisfying Clifford algebra $\{\G_A,\G_B\}=2\eta_{AB}$; the AdS generators $\hat{M}_{AB}$ are represented by $6\times 6$ \emph{super-matrices}, that reduce to $4\times 4$ matrices $M_{AB}=\frac{i}{4}[\Gamma_A,\Gamma_B]$ in the AdS subspace, see Appendix A. One choice of $\G$-matrices is $\Gamma_A =(i\gamma_a\gamma_5, \gamma_5)$, where $\g_a$ are the usual $4$D $\g$-matrices. In this particular representation, the components of $M_{AB}$ are given by $M_{ab}=\frac{i}{4}[\gamma_a,\gamma_b]=\frac12\sigma_{ab}$ and $M_{a5}=-\frac{1}{2}\gamma_a$. 
%%%%%%%%%%%%%%%%%%%%%%%%%%%%%%%%%%%%%%%%%%%%%%%%%%%%%%%

  The ten AdS bosonic generators $M_{AB}$ are accompanied by eight independent fermionic generators $\hat{Q}_{\a}^{I}$, with spinor index $\a=1,2,3,4$ and $SO(2)$ index $I=1,2$, comprising a pair of Majorana spinors, and one additional bosonic generator $\hat{T}$ related to $SO(2)\sim U(1)$ extension. Together, they satisfy $\mathfrak{osp}(4\vert 2)$ superalgebra (consistency requires that fermionic generators $\hat{Q}^{I}_{\a}$ transform as components of an AdS Majorana spinor): 
\begin{align}\label{OSp_algebra}
[\hat{M}_{AB},\hat{M}_{CD}]&=i(\eta_{AD}\hat{M}_{BC}+\eta_{BC}\hat{M}_{AD}-\eta_{AC}\hat{M}_{BD}-\eta_{BD}\hat{M}_{AC
})\ , \nn\\
[\hat{M}_{AB},\hat{Q}^{I}_{\a}]&=-(M_{AB})_{\a}^{\;\;\b}\hat{Q}^{I}_{\b}\ ,\nn\\
\{\hat{Q}^{I}_{\a},\hat{Q}^{J}_{\b}\}&=-2\delta^{IJ}(M^{AB}C^{-1})_{\a\b}\hat{M}_{AB}-i\varepsilon^{IJ}C_{\a\b}\hat{T}\ , \nn\\
[\hat{T},\hat{Q}^{I}_{\a}]&=-i\varepsilon^{IJ}\hat{Q}_{\a}^{I}\ ,
\end{align}
with antisymmetric tensor $\varepsilon^{IJ}$, $\varepsilon^{12}=1$. Matrix $C^{-1}$ is the inverse of the charge-conjugation matrix (spinor metric) for which we use the following representation given in terms of Pauli matrices: $C=-\s^{3}\otimes i\s^{2}$, and $C_{\a\b}=-C_{\b\a}$. Numerically we have $C^{-1}=-C$, but the index structure of the two is different since $C_{\a\g}(C^{-1})^{\g\b}=\delta_{\a}^{\b}$. More visually,   
\begin{equation}
\sbox0{$\begin{matrix}0&0\\0&0 \end{matrix}$}
\sbox1{$\begin{array}{c|c}0&-1\\ \hline 1&0 \end{array}$}
\sbox2{$\begin{array}{c|c}0&1\\ \hline -1&0 \end{array}$}
C=\left(
\begin{array}{c|c}
\usebox{1} & \makebox{$0_{2\times 2}$}  \\
\hline
 \makebox{$0_{2\times 2}$}   & \usebox{2}
\end{array}
\right)\ . \label{Cmatrix}
\end{equation} 
An explicit matrix representation of $\mathfrak{osp}(4|2)$ superalgebra is given in Appendix A. 

  By introducing a new set of (re-scaled) generators $\{\hat{\mathcal{M}}_{ab}:=\hat{M}_{ab},\hat{\mathcal{P}}_{a}:=\a\hat{M}_{a5}, \hat{\mathcal{Q}}_{\a}^{I}:=\sqrt{\a}\hat{Q}_{\a}^{I}, \hat{\mathcal{T}}:=\a\hat{T}\}$ we can recast the $\mathfrak{osp}(4\vert 2)$ superalgebra (\ref{OSp_algebra}) into the following form:
\begin{align}
[\hat{\mathcal{P}}_{a}, \hat{\mathcal{P}}_{b}]&=-i\a^{2}\hat{\mathcal{M}}_{ab}\ , \nn\\
[\hat{\mathcal{M}}_{ab}, \hat{\mathcal{P}}_{c}]&=i(\eta_{bc}\hat{\mathcal{P}}_{a}-\eta_{ac}\hat{\mathcal{P}}_{b})\ , \nn\\
[\hat{\mathcal{M}}_{ab}, \hat{\mathcal{M}}_{cd}]&=i(\eta_{ad}\hat{\mathcal{M}}_{bc}+\eta_{bc}\hat{\mathcal{M}}_{ad}-\eta_{ac}\hat{\mathcal{M}}_{bd}-\eta_{bd}\hat{\mathcal{M}}_{ac}) \ ,\nn\\ 
[\hat{\mathcal{P}}_{a},\hat{\mathcal{Q}}^{I}_{\a}]&=-\a(M_{a5})_{\a}^{\;\;\b}\hat{\mathcal{Q}}^{I}_{\b}\ ,\nn\\
[\hat{\mathcal{M}}_{ab},\hat{\mathcal{Q}}^{I}_{\a}]&=-(M_{ab})_{\a}^{\;\;\b}\hat{\mathcal{Q}}^{I}_{\b}\ , \nn\\
[\hat{\mathcal{T}},\hat{\mathcal{Q}}_{\a}^{I}]&=-i\varepsilon^{IJ}\hat{\mathcal{Q}}_{\a}^{I}\ , \nn\\
\{\hat{\mathcal{Q}}^{I}_{\a},\hat{\mathcal{Q}}^{J}_{\b}\}&=-2\delta^{IJ}\a(M^{ab}C^{-1})_{\a\b}\hat{\mathcal{M}}_{ab}-2\delta^{IJ}(M^{a5}C^{-1})_{\a\b}\hat{\mathcal{P}}_{a}-i\varepsilon^{IJ}C_{\a\b}\hat{\mathcal{T}}
\ . \label{OSp(4|2)-split}
\end{align}
Under WI contraction $\a\rightarrow 0$ it reduces to $N=2$ Poincar\'{e} superalgebra.

%%%%%%%%%%%%%%%%%%%%%%%%%%%%%%%%%%%%%%%%
  Orthosymplectic supergroup $OSp(2n\vert m)$ (symplectic sector is always even-dimensional) consists of those super-matrices $U$ that preserve the graded metric 
\begin{equation}
G=\left(\begin{array}{c|c}
\S_{\a\b} & 0_{2n\times m} \\
\hline
0_{m \times 2n} & \Delta_{ij}
\end{array}\right)\ ,
\end{equation}
with some real $2n\times 2n$ matrix $\S_{\a\b}=-\S_{\b\a}$, and some real $m\times m$ matrix $\Delta_{ij}=\Delta_{ji}$. Considering only infinitesimal transformations $U=1+\epsilon M$, generated by some $\mathfrak{osp}(2n\vert m)$-valued supermatrix 
\begin{equation}
M=\left(\begin{array}{c|c}
A & B \\
\hline
C & D
\end{array} \right)\ ,
\end{equation}
(bosonic blocks $A_{2n\times 2n}$ and $D_{m\times m}$ have ordinary \emph{commuting} entries, and fermionic blocks $B_{2n\times m}$ and $C_{m\times 2n}$ have \emph{Grassmann-valued} entries), the defining relation becomes
\begin{equation}
M^{ST}G+GM=0\ .
\end{equation} 
Super-transpose, super-hermitian adjoint and super-trace are defined by imposing the standard rules $(MN)^{ST}=N^{ST}M^{ST}$, $(MN)^{\dag}=N^{\dag}M^{\dag}$ and $STr(MN)=STr(NM)$,
\begin{equation}
M^{ST}=\left(\begin{array}{c|c}
A^{T} & C^{T} \\
\hline
-B^{T} & D^{T}
\end{array} \right),\;\;\; M^{\dag}=\left(\begin{array}{c|c}
A^{\dag} & C^{\dag} \\
\hline
B^{\dag} & D^{\dag}
\end{array} \right)\ ,\;\;\; \Str(M)=\tr(A)-\tr(D)\ .
\end{equation}

  Now, the key observation is that a pair of Majorana fields $\chi^{I}_{\m}$ (describing a pair of neutral spin-$3/2$ gravitini) constitute the fermionic sector of the
$\mathfrak{osp}(4\vert 2)$ connection super-matrix $\O_{\m}$. We can expand this super-connection over the basis $\{\hat{M}_{ab},\hat{M}_{a5},\hat{Q}_{\a}^{I},\hat{T}\}$ with the corresponding gauge fields $\{\o_{\m}^{\ ab},\o_{\m}^{\ a5},\bar{\chi}^{I}_{\m}, A_{\m}\}$, as
\begin{align}
\O_{\m}=\frac{1}{2}\o_{\m}^{\ ab}\hat{M}_{ab}+\o_{\m}^{\ a5}\hat{M}_{a5}+(\bar{\chi}_{\m}^{I})^{\a}\hat{Q}^{I}_{\a}+A_{\m}\hat{T}
=\sbox0{$\begin{array}{cc} \chi^{1}_{\m} & \chi^{2}_{\m}\end{array}$}
\sbox1{$\begin{array}{c} \bar{\chi}^{1}_{\m}\\ \bar{\chi}^{2}_{\m} \end{array}$}
\sbox2{$\begin{array}{c|c} 0& i A_{\m}\\ \hline -iA_{\m}&0 \end{array}$}
\left(
\begin{array}{c|cc}
\omega_{\m} & \usebox{0}  \\
\hline
\usebox{1}  & \usebox{2} 
\end{array}
\right)\ ,
\end{align} 
where we have $\mathfrak{so}(2,3)$ gauge field $\omega_{\m}=\tfrac{1}{2}\o_{\m}^{\ AB}M_{AB}=\tfrac{1}{2}\o_{\m}^{\ ab}M_{ab}+\o_{\m}^{\ a5}M_{a5}=\tfrac{1}{4}\o_{\m}^{\ ab}\s_{ab}-\tfrac{1}{2}\o_{\m}^{\ a5}\g_{a}$, a pair of Majorana vector-spinors $\chi^{I}_{\m}$ with components $(\chi^{I}_{\m})_{\a}$, and their Dirac-adjoints $\bar{\chi}^{I}_{\m}=-(\chi^{I}_{\m})^{T}C^{-1}$ with components $(\bar{\chi}^{I}_{\m})^{\a}=-(\chi^{I}_{\m})_{\b}(C^{-1})^{\b\a}$ ($\a=1,2,3,4$).

Equivalently, we can expand $\O_{\m}$ over the re-scaled basis $\{\hat{\mathcal{M}}_{ab},\hat{\mathcal{P}}_{a}, \hat{\mathcal{Q}}_{\a}^{I}, \hat{\mathcal{T}}\}$, but with a different set of gauge fields $\{\o_{\m}^{\ ab},e_{\m}^{a}:=\tfrac{1}{\a}\o_{\m}^{\ a5}, \bar{\psi}^{I}_{\m}:=\tfrac{1}{\sqrt{\a}}\bar{\chi}^{I}_{\m}, \mathcal{A}_{\m}:=\tfrac{1}{\a}A_{\m}\}$, as
\begin{align}
\O_{\m}=\frac{1}{2}\o_{\m}^{\ ab}\hat{\mathcal{M}}_{ab}+\tfrac{1}{\a}\o_{\m}^{\ a5}\hat{\mathcal{P}}_{a}+\tfrac{1}{\sqrt{\a}}\bar{\chi}_{\m}^{\a}\hat{\mathcal{Q}}_{\a}
=\sbox0{$\begin{array}{cc}\sqrt{\a}\psi^{1}_{\m} &\sqrt{\a}\psi^{2}_{\m}\end{array}$}
\sbox1{$\begin{array}{c}\sqrt{\a}\bar{\psi}^{1}_{\m}\\ \sqrt{\a}\bar{\psi}^{2}_{\m} \end{array}$}
\sbox2{$\begin{array}{c|c} 0& i\a \mathcal{A}_{\m}\\ \hline -i\a \mathcal{A}_{\m}&0 \end{array}$}
\left(
\begin{array}{c|cc}
\omega_{\m} & \usebox{0}  \\
\hline
\usebox{1}  & \usebox{2} 
\end{array}
\right)\ ,\label{Superconnection42}
\end{align} 
where we again have $\mathfrak{so}(2,3)$ gauge field $\o_{\m}=\frac{1}{4}\o_{\m}^{\ ab}\s_{ab}-\frac{\a}{2}e_{\m}^{a}\g_{a}$, two independent Majorana spinors $\psi^{I}_{\m}$, and (dimensionless) $U(1)$ vector potential $\mathcal{A}_{\m}$. We will use this particular representation because it makes WI contraction more transparent.

The two Majorana spinors, $\psi^{1}_{\m}$ and $\psi^{2}_{\m}$, can be combined into an $SO(2)$ doublet,
\begin{equation}
\Psi_{\m}=\begin{pmatrix}
  \psi^{1}_{\m}  \\
  \psi^{2}_{\m}
\end{pmatrix}\ .
\end{equation}
It can be readily confirmed that gauge supermatrix (\ref{Superconnection42}) satisfies the defining relation for the elements of $\mathfrak{osp}(4\vert 2)$ superalgebra ($C$ is the charge-conjugation matrix (\ref{Cmatrix})),
\begin{equation}
\sbox0{$\begin{array}{c|c} 1 & 0 \\ \hline 0 & 1 \end{array}$}
\sbox1{$\begin{array}{cc} 0 & 0 \end{array}$}
\sbox2{$\begin{array}{c} 0 \\ 0 \end{array}$}
\O_{\m}^{ST}\left(\begin{array}{c|cc}
  C & \usebox{1} \\
  \hline
  \usebox{2} & \usebox{0}   
\end{array}\right)+\left(\begin{array}{c|cc}
  C & \usebox{1} \\
  \hline
  \usebox{2} & \usebox{0}   
\end{array}\right)\O_{\m}=0\ .
\end{equation} 
%%%%%%%%%%%%%%%%%%%%%%%%%%%%%%%%%%%%%%%

\noindent
Field strength associated with AdS gauge field $\omega_\m$ is
\bea
F_{\mu\nu}=
\partial_\mu\omega_\nu-\partial_\nu\omega_\mu-i[\omega_\mu,\omega_\nu]=\Big( R_{\mu\nu}^{\ \ ab}-(\o_\mu^{\ a5}\o_\nu^{\ b5}-\o_\mu^{\ b5}\o_\nu^{\ a5})\Big)
\frac{\sigma_{ab}}{4}-F_{\mu\nu}^{\ \ a5}\frac{\gamma_a}{2}\ ,  \label{FabFa5} 
\eea
with (note that $D^{L}_{\m}$ stands for the Lorentz $SO(1,3)$ covariant derivative)
\begin{align}
R_{\m\n}^{\ \ ab}&=
\pa_\m\omega_\n^{\ ab}-\pa_\n\omega_\m^{\ ab}+\omega_{\m\ c}^{\ a}\;\omega_\n^{\ cb}
-\omega_{\n\ c}^{\ a}\;\omega_\m^{\ cb} \ , \label{Rab}\\
F_{\m\n}^{\ \ a5}&=D^{L}_\m \o^{\ a5}_\n-D^{L}_\n \o^{\ a5}_\m\ . \label{Ta}
\end{align}   

It was shown in the seventies that one can relate AdS gauge field theory to gravity (GR) by identifying $\o_{\m}^{\ ab}$ with the Lorentz spin-connection, $\o_{\m}^{\ a5}$ with the re-scaled vierbein field $\a e_{\m}^{a}$; vierbein is related to the metric tensor by $\eta_{ab}e_{\m}^{a}e_{\n}^{b}=g_{\m\n}$ and $e=\det(e_{\m}^{a})=\sqrt{-g}$. Consequently, $R_{\mu\nu}^{\ \ ab}$ can be identified with the curvature tensor, and $F_{\m\n}^{\ \ a5}$ with re-scaled torsion $\a T_{\mu\nu}^{\ \ a}$. Therefore, in the AdS setting, we have a natural unification of vierbein and spin-connection as components of a general $SO(2,3)$ gauge field; each transforms as a gauge field and stands on equal footing. In order to establish this identification, one has to break the original AdS gauge symmetry to the Lorentz $SO(1,3)$ gauge symmetry by introducing an auxiliary field $\phi=\phi^A\G_A$ \cite{stelle-west}. This field transforms in the adjoint representation of $SO(2,3)$ and it is constrained by $\eta_{AB}\phi^{A}\phi^{B}=l^{2}$. We can now start with an action of the Yang-Mills type, originally suggested by McDowell and Mansouri \cite{MacDowell-Mansouri}, invariant under $SO(2,3)$ gauge transformations:  
\begin{equation}\label{SAdS}
S_{AdS}=\frac{il}{64\pi G_N}\tr \int{\rm d}^4x\;\varepsilon^{\mu\nu\rho\sigma}
F_{\mu\nu} F_{\rho\sigma}\phi  \ ,
\end{equation}
where we have AdS covariant derivative in the adjoint representation,
\begin{equation}
D_\mu\phi = \partial_\mu\phi -i[\omega_\mu, \phi]\ . 
\end{equation}
We choose the \emph{physical gauge} by setting $\phi^a=0$ and $\phi^5=l$, and thus obtain:
\begin{equation} 
S_{AdS}|_{\text{g.f.}}=-\frac{1}{16\pi G_{N}}\int 
\diff^{4}x \Bigg(e\big(R(e,\o) -6/l^2\big)+\frac{l^2}{16}R_{\m\n}^{\;\;\ ab}R_{\r\s}^{\;\;\ cd}\varepsilon^{\m\n\r\s}
\varepsilon_{abcd}
\Bigg)\ , 
\end{equation}
which is the standard GR action (written in the first order formalism) involving the Einstein-Hilbert term, negative cosmological constant $\L=-3/l^{2}=-3\a^{2}$, and the topological Gauss-Bonet term that can be omitted. 

\noindent
Therefore, we can write the $SO(2,3)$ field strength as
\bea
F_{\mu\nu}=\frac{1}{4}\Big( R_{\mu\nu}^{\ \ ab}-\a^{2}(e_\mu^ae_\nu^b-e_\mu^be_\nu^a)\Big)\sigma_{ab}- \frac{\a}{2}T_{\mu\nu}^{\ \ a}\gamma_a\ ,  
\eea
and we see that the vierbein and torsion terms vanish under WI contraction.

By generalization, we introduce ${Osp}(4\vert 2)$ field strength $\mathbb{F}_{\m\n}$ associated with the super-connection $\O_{\m}$,
\begin{align}
\mathbb{F}_{\m\n}&=\partial_{\m}\Omega_{\n}-\partial_{\n}\Omega_{\m}-i[\Omega_{\m},\Omega_{\n}]
\nn \\
&=\sbox0{$\begin{array}{c|c} 0 & i\a\widetilde{\mathcal{F}}_{\m\n} \\ \hline -i\a\widetilde{\mathcal{F}}_{\m\n} & 0 \end{array}$}
\sbox1{$\begin{array}{cc} \sqrt{\a}(\mathcal{D}_{\m}\psi^{1}_{\n}-\mathcal{D}_{\n}\psi^{1}_{\m}) & \sqrt{\a}(\mathcal{D}_{\m}\psi^{2}_{\n}-\mathcal{D}_{\n}\psi^{2}_{\m}) \end{array}$}
\sbox2{$\begin{array}{c}\sqrt{\a}(\mathcal{D}_{\m}\bar{\psi}^{1}_{\n}-\mathcal{D}_{\n}\bar{\psi}^{1}_{\m}) \\  \sqrt{\a}(\mathcal{D}_{\m}\bar{\psi}^{2}_{\n}-\mathcal{D}_{\n}\bar{\psi}^{2}_{\m}) \end{array}$}
\left(\begin{array}{c|cc}
  \widetilde{F}_{\m\n} & \usebox{1}   \\
  \hline
  \usebox{2} & \usebox{0}
\end{array}\right)\ , \label{SFstreangth(42)}
\end{align}
with extended AdS field strength $\widetilde{F}_{\m\n}$ (summation over $I=1,2$ is implied)
\begin{align}
\widetilde{F}_{\m\n}&=F_{\m\n}-i\a(\psi^{I}_{\m}\bar{\psi}^{I}_{\n}-\psi^{I}_{\n}\bar{\psi}^{I}_{\m})=\frac{1}{4}\widetilde{R}_{\m\n}^{\ \ mn}\s_{mn}-\frac{\a}{2}\widetilde{T}_{\m\n}^{\ \ m}\g_{m}\ ,
\end{align}
involving extended curvature tensor $\widetilde{R}_{\m\n}^{\ \ mn}$ and extended torsion $\widetilde{T}_{\m\n}^{\ \ m}$, given by 
\begin{align}\label{SR}
\widetilde{R}_{\m\n}^{\ \ mn}&:=R_{\m\n}^{\ \ mn}-\a^{2}(e_{\m}^{m}e_{\n}^{n}-e_{\m}^{n}e_{\n}^{m})-i\a(\bar{\Psi}_{\m}\s^{mn}\Psi_{\n})\ , \\ 
\widetilde{T}_{\m\n}^{\ \ m}&:=T_{\m\n}^{\ \ m}+i(\bar{\Psi}_{\m}\g^{m}\Psi_{\n})\ .\label{ST}
\end{align}
Electromagnetic field strength is also modified by a bilinear current term $\mathcal{J}_{(e)}$,
\begin{equation}
\widetilde{\mathcal{F}}_{\m\n}:=\mathcal{F}_{\m\n}-\mathcal{J}_{(e)\m\n}=\partial_{\m}\mathcal{A}_{\n}-\partial_{\n}\mathcal{A}_{\m}-\bar{\Psi}_{\m}i\s^{2}\Psi_{\n}\ .\label{U(1)_strength}
\end{equation}
Note that Pauli matrix $i\s^{2}$ mixes the two Majorana components in $\mathcal{J}_{(e)}$. 

\noindent
In the fermionic sector of $\mathbb{F}_{\m\n}$ we introduced 
\begin{align}
\mathcal{D}_{\m}\psi^{1}_{\n}&:=D_{\m}\psi^{1}_{\n}+\a \mathcal{A}_{\m}\psi^{2}_{\n}\ , \label{Dpsi1}\\
\mathcal{D}_{\m}\psi^{2}_{\n}&:=D_{\m}\psi^{2}_{\n}-\a \mathcal{A}_{\m}\psi^{1}_{\n}\ , \label{Dpsi2}
\end{align}
where $D_{\m}$ stands for $SO(2,3)$ covariant derivative. The fact that Majorana spinors $\psi^{1}_{\m}$ and $\psi^{2}_{\n}$ are not charged is reflected in the manner in which they couple to the gauge field $\mathcal{A}_{\m}$. Using them, we can define two charged Dirac vector-spinors $\psi^{\pm}_{\m}=\psi^{1}_{\m}\pm i\psi^{2}_{\m}$, related to each other by $C$-conjugation, $\psi^{-}_{\m}=\psi^{c+}_{\m}=C\bar{\psi}^{+ T}_{\m}$, that do couple to $\mathcal{A}_{\m}$ in the right way. Using the Pauli matrix $i\s^{2}$ we can unify (\ref{Dpsi1}) and (\ref{Dpsi2}) as
\begin{equation}
\mathcal{D}_{\m}\Psi_{\n}=(D_{\m}+\a \mathcal{A}_{\m}i\s^{2})\Psi_{\n}=\left(D^{L}_{\m}
+\frac{i\a}{2}\g_{\m}
+\a \mathcal{A}_{\m}i\s^{2}\right)\Psi_{\n}\ .
\end{equation}
%%%%%%%%%%%%%%%%%%%%%%%%%%%%%%%%%%%%%%%%%%%%%%

  Now consider an action, similar to the one defined in (\ref{SAdS}) for pure gravity, but now appropriately generalized to be invariant under extended $OSp(4\vert 2)$ gauge transformations,
\begin{equation}
S_{42}=\Str\int{\rm d}^4x \;\varepsilon^{\mu\nu\rho\sigma}
\mathbb{F}_{\mu\nu}(a\mathbb{I}_{6\times 6}+b\Phi^{2}/l^{2})\mathbb{F}_{\rho\sigma}\Phi  \ .\label{KomDejstvo(42)} 
\end{equation} 
The action is real and we introduced a pair of free parameters, $a$ and $b$, that will by fixed later. The first part of (\ref{KomDejstvo(42)}) is the quadratic Yang-Mills type of action, and the second part ($b$-term) is necessary for having local SUSY after the symmetry braking.  

Generalized auxiliary field $\Phi$ is given by a supermatrix (we now have two Majorana spinors $\l_{1}$ and $\l_{2}$, and additional scalar fields $\pi$, $m$ and $\s$), see also \cite{Cham1, Cham2, Cham3, NCSUGRA2}, 
\begin{equation}
\sbox0{$\begin{array}{c|c} \pi-\s & m \\ \hline m & \s \end{array}$}
\sbox1{$\begin{array}{cc} \l_{1} & \l_{2} \end{array}$}
\sbox2{$\begin{array}{c} -\bar{\l}_{1} \\ -\bar{\l}_{2} \end{array}$}
\Phi=\left(\begin{array}{c|cc}
  \tfrac{1}{4}\pi+i\phi^{a}\g_{a}\g_{5}+\phi^{5}\g_{5} & \usebox{1} \\
  \hline
   \usebox{2} & \usebox{0} \\
\end{array}\right)\ .
\end{equation}
In the physical gauge $\l_{1}=\l_{2}=\pi=\s=m=\phi^{a}=0$, and $\phi^{5}=l$, yielding
\begin{equation}\label{Phi_SB}
\Phi\vert_{\text{g.f.}}=
\sbox0{$\begin{array}{c|c} 0 & 0 \\ \hline 0 & 0 \end{array}$}
\sbox1{$\begin{array}{cc} 0 & 0 \end{array}$}
\sbox2{$\begin{array}{c} 0 \\ 0 \end{array}$}
\left(\begin{array}{c|cc}  
l\g_{5} & \usebox{1} \\
  \hline
\usebox{2} & \usebox{0} \\  
\end{array}\right)\ .
\end{equation}

Field strength $\mathbb{F}_{\m\n}$ and the auxiliary field transform in the adjoint representation of $OSp(4|2)$, with infinitesimal variations
\begin{align}\label{42var}
\delta_{\epsilon}\mathbb{F}_{\m\n}=i[\epsilon,\mathbb{F}_{\m\n}]\ , \;\;\;
\delta_{\epsilon}\Phi=i[\epsilon,\Phi]\ ,
\end{align}
for some $\mathfrak{osp}(4\vert 2)$-valued gauge parameter $\epsilon$ given by a supermatrix
\begin{equation}
\epsilon=
\sbox0{$\begin{array}{c|c} 0 & i\a \\ \hline -i\a & 0 \end{array}$}
\sbox1{$\begin{array}{cc} \xi_{1} & \xi_{2} \end{array}$}
\sbox2{$\begin{array}{c} \bar{\xi}_{1} \\ \bar{\xi}_{2} \end{array}$}
\left(\begin{array}{c|cc}
\frac{1}{2}\epsilon^{AB}M_{AB} & \usebox{1}\\
\hline
\usebox{2}  & \usebox{0}
 \end{array}
\right)\ .
\end{equation}
From (\ref{42var}) the invariance of the action (\ref{KomDejstvo(42)}) follows immediately.
 
  After the gauge fixing, field $\Phi^{2}/l^{2}$ that appears in the second term of (\ref{KomDejstvo(42)}) becomes a \emph{projector} that reduces any $\mathfrak{osp}(4\vert 2)$ supermatrix to its $\mathfrak{so}(2,3)$ sector, and the classical $OSp(4|2)$ gauge-invariant action (\ref{KomDejstvo(42)}) reduces to
\begin{equation}\label{S42-split}
S_{42}\vert_{\text{g.f.}}=\int\diff^{4}x\;\varepsilon^{\m\n\r\s}\Bigg(\frac{(a+b)il}{4}\widetilde{R}_{\m\n}^{\ \ mn}\widetilde{R}_{\r\s}^{\ \ rs}\varepsilon_{mnrs}-4al(\mathcal{D}_{\m}\bar{\Psi}_{\n}\g_{5}\mathcal{D}_{\r}\Psi_{\s})\Bigg)\ .
\end{equation}
The term that is quadratic in the Lorentz $SO(1,3)$ covariant derivative $D_{\m}^{L}$ can be transformed by partial integration,
\begin{equation}
\int\diff^{4}x\;\varepsilon^{\m\n\r\s}(D^{L}_{\m}\bar{\Psi}_{\n}\g_{5}D^{L}_{\r}\Psi_{\s})=\frac{1}{16}\int\diff^{4}x\;\varepsilon^{\m\n\r\s} 
R_{\m\n}^{\ \ mn}(\bar{\Psi}_{\r}\s^{rs}\Psi_{\s})\varepsilon_{mnrs}\ , \label{Term}
\end{equation}
where we invoked the commutator of two Lorentz covariant derivatives
\begin{equation}
i[D^{L}_{\m},D^{L}_{\n}]\Psi_{\s}=\tfrac{1}{4}R_{\m\n}^{\ \ mn}\s_{mn}\Psi_{\s}\ .
\end{equation}
Term of the same type appears in the first part of the action (\ref{S42-split}). These two contributions have to cancel each other in order to have SUSY, and this implies the constraint $b=-a/2$. Moreover, to obtain the correct normalization of the Einstein-Hilbert term, we set $a=il/32\pi G_{N}=il/4\k^{2}$, yielding 
\begin{align} \label{FinalAction(42)}
&S_{42}\vert_{\text{g.f.}}=-\frac{1}{2\k^{2}}\int \diff^{4}x\;\Bigg(e\Big(R(e,\o)-6\a^{2}\Big)+\frac{1}{16\a^{2}}R_{\m\n}^{\ \ mn}R_{\r\s}^{\ \ rs}\varepsilon^{\m\n\r\s}\varepsilon_{mnrs}\\
&+\varepsilon^{\m\n\r\s}\Big(2\bar{\Psi}_{\m}\g_{5}\g_{\n}(D_{\r}+\a \mathcal{A}_{\r}i\s^{2})\Psi_{\s}+i\mathcal{F}_{\m\n}(\bar{\Psi}_{\r}\g_{5}i\s^{2}\Psi_{\s})-\frac{i}{2}(\bar{\Psi}_{\m}i\s^{2}\Psi_{\n})(\bar{\Psi}_{\r}\g_{5}i\s^{2}\Psi_{\s})\Big)\Bigg) \nn \ . 
\end{align}
%\nn \\=&-\frac{1}{16\pi G_{N}}\int \diff^{4}x\;\Bigg(e\Big(R-6\a^{2}\Big)+\frac{1}{16\a^{2}}R_{\m\n}^{mn}R_{\r\s}^{rs}\varepsilon^{\m\n\r\s}\varepsilon_{mnrs}\Bigg)\nn \\&+2\varepsilon^{\m\n\r\s}\bar{\Psi}_{\m}\g_{5}\g_{\n}(D_{\r}+\a A_{\r}i\s^{2})\Psi_{\s}+2\mathcal{F}\cdot\mathcal{J}_{(m)}-\mathcal{J}_{(e)}\cdot\mathcal{J}_{(m)}

  However, this is \emph{not} the full $N=2$ $AdS_{4}$ SUGRA action. The gravity part is correct (we can omit the topological Gauss-Bonet term) and we also get the correct kinetic term for the gravitino doublet. There are also two bilinear source terms, electric and magnetic, 
\begin{align}
\mathcal{J}_{(e)\m\n}:=\bar{\Psi}_{\m}i\s^{2}\Psi_{\n}\ ,\;\;\;
\mathcal{J}_{(m)}^{\m\n}:=\tfrac{i}{2e}\varepsilon^{\m\n\r\s}(\bar{\Psi}_{\r}\g_{5}i\s^{2}\Psi_{\s})\ .
\end{align}
But we are missing the contribution from the $SO(2)$ part of the bosonic sector, in particular, the kinetic term for $U(1)$ gauge field $\mathcal{A}_{\m}$. The reason for this defect can be traced back to the specific form that the auxiliary field assumes in the physical gauge $\Phi|_{\text{g.f.}}$ (\ref{Phi_SB}); it completely annihilates the $SO(2)$ sector of any $\mathfrak{osp}(4\vert 2)$ supermatrix. To restore the missing terms, we must introduce an additional action, supplementing (\ref{KomDejstvo(42)}). In \cite{U(1)}, following the approach of \cite{PLgaugefield}, we defined a classical action invariant under $SO(2,3)\times U(1)$ gauge transformations ($\sim$bosonic sector of $OSp(4\vert 2)$) that involves an \emph{additional auxiliary field} $f=\frac{1}{2}f^{AB}M_{AB}$. Its role is to produce the canonical kinetic term for $U(1)$ gauge field in the absence of Hodge dual operator (this is of course the crucial point, we are trying to construct a purely geometrical action that does not involve the metric tensor $g_{\m\n}$ explicitly). This auxiliary field $f$ is a $U(1)$-neutral $0$-form that takes values in $\mathfrak{so}(2,3)$ algebra, and it transforms in the adjoint representation of $SO(2,3)$.

%The $SO(2,3)\times U(1)$ gauge field is simply  \bev_{\m}=\omega_{\m}+A_{\m}\ .\label{MasterGaugePot}\eeThe corresponding field strength\be f_{\m\n}=\partial_\m v_\n-\partial_\n v_\m-i[v_\m,v_\n]=F_{\m\n}+\mathcal{F}_{\m\n}\ ,\ee splits into $SO(2,3)$ field strength $F_{\m\n}$ and electromagnetic field strength $\mathcal{F}_{\m\n}$.\noindent The action before gauge fixing is \bea S_{A}= -\frac{\a}{16} \tr\int {\rm d}^{4}x\;\varepsilon^{\mu\nu\rho\sigma}\Bigg( f  f_{\m\n}D_\r\phi D_\s\phi\phi+\frac{i}{3!}f^{2}D_\m\phi D_\n\phi D_\r\phi D_\s\phi\phi \Bigg) + c.c. \label{ActionGaugeComm}\eea After calculating traces and symmetry braking we obtain \begin{align} S_{A}|_{\text{g.f.}}& = \frac{1}{2}\int {\rm d}^{4}x\;e\; \Big( f^{ab}e_a^\m e_b^\n\mathcal{F}_{\m\n} +\frac{1}{2}(f^{ab}f_{ab}+2f^{a5}f_{a}^{\;\;5})\Big) \ . \label{SA} \end{align}
%%%%%%%%%%%%%%%%%%%%%%%%%%%%%%%%%%%% \noindent Equations of motion (EoMs) for the components of the auxiliary field $f$ are \begin{align}f_{a5}=0\ , \;\;\;\;f_{ab}=-e_a^\m e_b^\n\mathcal{F_{\m\n}}\ . \label{EOMs-0th order}\end{align}Using these EoMs we can eliminate the auxiliary field in the action (\ref{SA}). This leaves us with the canonical kinetic term for pure $U(1)$ gauge field in curved space-time\be\label{EM_kin_term}S_{A}|_{\text{g.f.}}= -\frac{1}{4}\int {\rm d}^{4}x\;e\;g^{\mu\rho}g^{\nu\sigma}\mathcal{F}_{\m\n}\mathcal{F}_{\rho\sigma}\ .\ee 

The way to proceed is to employ this auxiliary field method to include the modified $U(1)$ field strength $\widetilde{\mathcal{F}}_{\m\n}$ defined in (\ref{U(1)_strength}). However, there seems to be no way to construct an $OSp(4\vert 2)$ gauge invariant action that is compatible with this procedure. Therefore, we will use an action, analogues to the one in \cite{U(1)}, invariant under purely bosonic $SO(2,3)\times U(1)$ sector of $OSp(4\vert 2)$, involving the bosonic field strength $\widetilde{f}_{\m\n}:=\widetilde{F}_{\m\n}+\kappa^{-1}\widetilde{\mathcal{F}}_{\m\n}=\widetilde{F}_{\m\n}+\kappa^{-1}(\mathcal{F}_{\m\n}-\mathcal{J}_{(e)\m\n})$ of $SO(2,3)\times U(1)$. The action is given by 
\begin{equation} \label{SA_SUGRA}
S_{A}=\tr\int{\rm d}^{4}x\;\varepsilon^{\mu\nu\rho\sigma}
\Bigg(c\;f \widetilde{f}_{\m\n}D_\r\phi
D_\s\phi\phi + d\;f^{2}D_\m\phi D_\n\phi D_\r\phi D_\s\phi\phi \Bigg) + c.c. 
\end{equation}
Note that, by doing this, we lose the complete $OSp(4|2)$ gauge invariance of the undeformed action before the symmetry breaking. Nevertheless, we will obtain the correct action for $N=2$ $SdS_{4}$ SUGRA in the physical gauge, and this is the only requirement that has to be satisfied in order to perform NC deformation.  

\noindent After calculating traces (see Appendix B) we obtain
\begin{align}
S_{A}= &\int {\rm d}^{4}x\;
\varepsilon^{\mu\nu\rho\sigma}\Bigg(\frac{ic}{2} f^{AB}F_{\m\n}^{\ \ CD}(D_\r\phi)^{E}(D_\s\phi)^{F}\phi^{G
}(\eta_{FG}\varepsilon_{ABCDE}
+2\eta_{AD}\varepsilon_{BCEFG}) \nn\\
&+c\kappa^{-1}f^{AB}\widetilde{\mathcal{F}}_{\m\n}(D_\r\phi)^{E}(D_\s\phi)^{F}
\phi^{G}\varepsilon_{ABEFG} \nn\\
&-\frac{id}{2}f^{AB}f_{AB}(D_\m\phi)^{E}(D_\n\phi)^{F}(D_\r\phi)^{G}
(D_\s\phi)^{H}\phi^{R}\varepsilon_{EFGHR} \Bigg) +c.c.  
\end{align}
We conclude that parameter $c$ must be real, otherwise the second term, involving $\widetilde{\mathcal{F}}_{\m\n}$, would be purely imaginary and would not contribute (and this term is the one that we need to include). Therefore, assuming real $c$, the first term (involving gravitational quantities like curvature tensor and torsion) becomes purely imaginary and vanishes after adding its complex conjugate (c.c.). Also, $d$ must be purely imaginary for the procedure to work. 

\noindent Gauge fixing yields
\begin{align}\label{SAgf}
S_{A}|_{\text{g.f.}}=\int {\rm d}^{4}x\;e
\Big(-8lc\kappa^{-1} f^{ab}\widetilde{\mathcal{F}}_{\m\n}e_{\m}^{a}e_{\n}^{b}+24ild\;f^{AB}f_{AB}\Big)\ . 
\end{align}
By varying this gauge fixed action over $f^{ab}$ and $f^{a5}$ independently, we obtain algebraic equations of motion (EoMs) for the components $f_{ab}$ and $f_{a5}$ of the auxiliary field $f$, respectfully, and they are given by
\begin{equation}\label{EoMs}
f_{ab}=-\frac{ic}{6\kappa d}\widetilde{\mathcal{F}}_{\m\n}e^{\m}_{a}e^{\n}_{b}\ ,\;\;\;f_{a5}=0\ .
\end{equation}
Inserting them back into the action (\ref{SAgf}), we obtain
\begin{align}\label{EM_kin_term}
S_{A}|_{\text{g.f.}}&=\frac{2ilc^{2}}{3\kappa^{2}d}\int{\rm d}^{4}x\;e\;\widetilde{\mathcal{F}}^{2}\ .
\end{align}
%%%%%%%%%%%%%%%%%%%%%%%%%%%%%%%%%%%%%%
To get the consistent normalization we set the prefactor to $(8\k^{2})^{-1}$, yielding another constraint $16ilc^{2}=3d$ for the parameters $c$ and $d$. To make the connection with the results of \cite{U(1)}, we take $c=1/32l$ and $d=i/192 l$, implying 
\begin{equation}
f_{ab}=-\kappa^{-1}\widetilde{\mathcal{F}}_{\m\n}e^{\m}_{a}e^{\n}_{b}\ .
\end{equation}
Therefore, after imposing the physical gauge, the original bosonic action (\ref{SA_SUGRA}), invariant under $SO(2,3)\times U(1)$ gauge transformations, reduces to $SO(1,3)\times U(1)$ gauge-invariant action containing the canonical kinetic term for $U(1)$ gauge field $\mathcal{A}_{\m}$ in curved space-time, and two additional terms involving gravitino current $\mathcal{J}_{(e)\m\n}=\bar{\Psi}_{\m}i\s^{2}\Psi_{\n}$, 
\begin{equation}
S_{A}|_{\text{g.f.}}=\frac{1}{4\k^{2}}\int{\rm d}^{4}x\;e\;\widetilde{\mathcal{F}}^{2}=\frac{1}{4\k^{2}}\int {\rm d}^{4}x\;e\;\left(\mathcal{F}^{2}-2\mathcal{F}\cdot\mathcal{J}_{(e)}+\mathcal{J}_{(e)}^{2}\right)\ .
\end{equation}

This is exactly the piece that was missing in (\ref{FinalAction(42)}). With this result in hand, we have the complete classical $N=2$ $AdS_{4}$ SUGRA action \cite{Ortin, Freedman}, 
\begin{align} \label{Final42classicalMajorana}
(S_{42}+S_{A})|_{\text{g.f.}}=-&\frac{1}{2\k^{2}}\int\diff^{4}x\;e\;\Bigg(R-6\a^{2}+2e^{-1}\varepsilon^{\m\n\r\s}\bar{\Psi}_{\m}\g_{5}\g_{\n}(D_{\r}+\a\mathcal{A}_{\r}i\s^{2})\Psi_{\s}\nn\\
&+2\mathcal{F}\cdot\mathcal{J}_{(m)}-\mathcal{J}_{(e)}\cdot\mathcal{J}_{(m)} 
-\frac{1}{4}\left(\mathcal{F}^{2}+\mathcal{J}_{(e)}^{2}-2\mathcal{F}\cdot\mathcal{J}_{(e)}\right)\Bigg)  \ .
\end{align}
The most important characteristics of this SUGRA model are the negative cosmological constant $\L=-3\a^{2}=-3/l^{2}$, and the fact that $U(1)$ coupling strength is equal to the WI contraction parameter $\a$. Under WI contraction ($\alpha\rightarrow 0$) the $N=2$ $AdS_{4}$ SUGRA action consistently reduces to the $N=2$ Poincar\'{e} SUGRA action.

In terms of charged Dirac vector-spinors $\psi^{\pm}_{\m}=\psi^{1}_{\m}\pm i\psi^{2}_{\m}$ (actually, we can use only one of them since they are related to each other by $C$-conjugation) the action becomes
\begin{align} \label{Final42classicalDirac}
(S_{42}+S_{A})|_{\text{g.f.}}=-&\frac{1}{2\k^{2}}\int\diff^{4}x\;e\Bigg(R(e,\o)-6\a^{2}
+2e^{-1}\varepsilon^{\m\n\r\s}\bar{\psi}^{+}_{\m}\g_{5}\g_{\n}(D_{\r}-i\a \mathcal{A}_{\r})\psi^{+}_{\s} \nn\\
&+2\mathcal{F}\cdot\mathcal{J}^{+}_{(m)}-\mathcal{J}^{+}_{(e)}\cdot\mathcal{J}^{+}_{(m)}-\frac{1}{4}\left(\mathcal{F}^{2}-2\mathcal{F}\cdot\mathcal{J}^{+}_{(e)}+(\mathcal{J}^{+}_{(e)})^{2}\right)\Bigg)\ ,
\end{align}
with $\mathcal{J}^{+}_{(e)}=\frac{1}{2i}(\bar{\psi}_{\m}^{+}\psi_{\n}^{+}-\bar{\psi}_{\n}^{+}\psi_{\m}^{+})$ and $\mathcal{J}^{+}_{(m)}=\frac{1}{4e}(\bar{\psi}_{\m}^{+}\g_{5}\psi_{\n}^{+}-\bar{\psi}_{\n}^{+}\g_{5}\psi_{\m}^{+})$.

For later purpose, we note that action (\ref{Final42classicalDirac}) contains a mass-like term for charged gravitino (we absorb the parameter $\kappa^{-1}$ into $\psi_{\m}^{+}$ to obtain the canonical dimensions),
\begin{equation}\label{mass-like term}
i\a\int\diff^{4}x\;e\;\bar{\psi}_{\m}^{+}\s^{\m\n}\psi^{+}_{\n}\ ,
\end{equation}
with mass-like parameter equal to the WI contraction parameter.

%%%%%%%%%%%%%%%%%%%%%%%%%%%%%%%%%%%%%%%%%
\subsection{$OSp(4\vert 1)$ SUGRA}
%%%%%%%%%%%%%%%%%%%%%%%%%%%%%%%%%%%%%%%%%
The $OSp(4\vert 1)$ supergroup has $14$ generators; ten bosonic AdS generators $\hat{M}_{AB}$, and  four fermionic generators, $\hat{Q}_{\a}$, comprising a single Majorana spinor (describing a single neutral gravitino). 
Supermatrix for the $OSp(4\vert 1)$ gauge field $\O_{\m}$ is given by
\begin{equation}
\sbox0{$\begin{matrix} 0&0\\ 0&0 \end{matrix}$}
\sbox1{$\begin{matrix} 0&-1\\ 1&0 \end{matrix}$}\sbox2{$\begin{matrix}0&1\\-1&0 \end{matrix}$}
\O_{\m}=\left(\begin{array}{c|c} \o_{\m} & \sqrt{\a}\psi_{\m} \\ \hline \sqrt{\a}\bar{\psi}_{\m}  & 0\end{array}\right)\ .
\end{equation} 
%%%%%%%%%%%%%%%%%%%%%%%%%%%%%%%%%%%%%%%
\noindent Consider the following action invariant under $OSp(4\vert 1)$ gauge transformations \cite{NCSUGRA1},
\begin{equation}\label{KomDejstvo(41)} 
S_{41}=\frac{il}{32\pi G_N}\;\Str\int{\rm d}^4x \;\varepsilon^{\mu\nu\rho\sigma}
\mathbb{F}_{\mu\nu}(\mathbb{I}_{5\times 5}-\tfrac{1}{2l^{2}}\Phi^{2})\mathbb{F}_{\rho\sigma}\Phi  \ .
\end{equation}
Auxiliary field is 
\begin{equation}
\Phi=\left(
\begin{array}{c|c}
\tfrac{1}{4}\pi+i\phi^{a}\g_{a}\g_{5}+\phi^{5}\g_{5} & \l \\
\hline
 -\bar{\l} & \pi
\end{array}
\right)\myeqe\left(
\begin{array}{c|c}
l\g_{5} & 0 \\
\hline
 0 & 0
\end{array}
\right)\ . 
\end{equation}
In the physical gauge, the $OSp(4\vert 1)$ gauge-invariant action (\ref{KomDejstvo(41)}) \emph{exactly} reduces to $N=1$ $AdS_{4}$ SUGRA action \cite{Ortin, Freedman, NCSUGRA2},
\begin{align} \label{FinalActionS1}
S_{41}|_{\text{g.f.}}&=-\frac{1}{2\kappa^{2}}\int \diff^{4}x\;\Bigg(e\big(R(e,\o)-6/l^{2}\big)+2\varepsilon^{\m\n\r\s}(\bar{\psi}_{\m}\g_{5}\g_{\n}D_{\r}\psi_{\s})\Bigg)\\
&=-\frac{1}{2\kappa^{2}}\int \diff^{4}x\;e\Bigg(R(e,\o)-6\a^{2}+
2e^{-1}\varepsilon^{\m\n\r\s}(\bar{\psi}_{\m}\g_{5}\g_{\n}D^{L}_{\r}\psi_{\s})-2i\a(\bar{\psi}_{\m}\s^{\m\n}\psi_{\n})\Bigg)\ . \nn
\end{align}
It contains Einstein-Hilbert term with the negative cosmological constant $\L=-3/l^{2}$, Rarita-Schwinger kinetic term for \emph{neutral} gravitino, and a mass-like gravitino term that is
needed in the presence of the cosmological constant to insure the invariance under local SUSY (gravitino actually remains massless). Topological Gauss-Bonet term is omitted. Cosmological constant and the mass-like term vanish under WI contraction, yielding minimal $N=1$ Poincar\'{e} SUGRA. Note that we do not need additional action terms in (\ref{KomDejstvo(41)}) to obtain a consistent classical theory. 

It is shown in \cite{NCSUGRA2} that linear (in $\theta^{\m\n}$) NC correction to (\ref{KomDejstvo(41)}) vanishes, and that one has to calculate the second order NC correction in order to see NC effects, which is exceedingly difficult. In the following section, we use the Seiberg-Witten approach to NC gauge field theories, to calculate linear NC correction to $N=2$ $AdS_{4}$ SUGRA, and conclude that \emph{it is not equal to zero}. The non-vanishing part comes from the additional bosonic action, $S_{A}$.
%%%%%%%%%%%%%%%%%%%%%%%%%%%%%%%%%%%%%%%%%%%%%%%%%%%%%%%%%%%%%%%%%%%%%%%%

\initiate 
%%%%%%%%%%%%%%%%%%%%%%%%%%%%%%%%%%%%%%%%%%%%%%%
\section{NC deformation}
%%%%%%%%%%%%%%%%%%%%%%%%%%%%%%%%%%%%%%%%%%%%%%%
Canonical deformation of the orthoymplectic action (\ref{KomDejstvo(42)}) is obtained by replacing ordinary commutative field multiplication with Moyal $\star$-product, yielding an NC action (denoted by ``$\star$'') manifestly invariant under NC-deformed $OSp(4\vert 2)_{\star}$ gauge transformations,
\begin{equation}
S^{\star}_{42}=\frac{il}{32\pi G_N}\Str\;\int{\rm d}^4x\;\varepsilon^{\mu\nu\rho\sigma}\left(
\hat{\mathbb{F}}_{\mu\nu}\star\hat{\mathbb{F}}_{\rho\sigma}\star\hat{\Phi}-\tfrac{1}{2l^{2}}\hat{\mathbb{F}}_{\mu\nu}\star\hat{\Phi}\star\hat{\Phi}\star\hat{\mathbb{F}}_{\rho\sigma}\star\hat{\Phi}\right)\ .  \label{NC42}
\end{equation}
Likewise, we have canonically deformed version of the bosonic action (\ref{SA_SUGRA}) with $c=1/32l$ and $d=i/192 l$,
\begin{align}\label{NCSA}
S^{\star}_{A}=\frac{1}{32l}\tr\int{\rm d}^{4}x\;\varepsilon^{\mu\nu\rho\sigma}
&\Big(\hat{f}\star\hat{\widetilde{f}}_{\m\n}\star D_\r\hat{\phi}
\star D_\s\hat{\phi}\star\hat{\phi}\nn\\
&+\frac{i}{6}\hat{f}\star\hat{f}\star D_\m\hat{\phi}\star D_\n\hat{\phi}\star D_\r\hat{\phi}\star D_\s\hat{\phi}\star\hat{\phi}\Big) + c.c.
\end{align} 
We denote NC fields by a ``hat'' symbol. 

In the Seiberg-Witten approach \cite{NCbook, SWMap1, SWMap2, SWMap3, SWMap4}, NC gauge field theory is completely defined by its commutative (classical) counterpart. For some non-Abelian gauge group $\mathcal{G}$ with generators $T_{A}$ satisfying Lie algebra relations $[T_{A},T_{B}]=if_{A\ \;B}^{\;\ C}T_{C}$, commutator of two infinitesimal gauge transformations $\delta_{\epsilon_{1}}$ and $\delta_{\epsilon_{2}}$ closes in the algebra,
\begin{equation}
[\delta_{\epsilon_{1}},\delta_{\epsilon_{2}}]=\delta_{-i[\epsilon_{1},\epsilon_{2}]}\ . 
\end{equation}
There is, however, a difficulty, in general, concerning the closure axiom for NC gauge transformations. Namely, for a given pair of NC gauge parameters $\hat{\L}_{1}$ and $\hat{\L}_{2}$ we would like to find a third one, $\hat{\L}_{3}$, such that 
\be 
[\delta^{\star}_{1}\ds\delta^{\star}_{2}]=\delta^{\star}_{3}\ .
\ee
Now, if NC gauge parameter $\hat{\L}$ is supposed to be Lie algebra-valued, $\hat{\Lambda}(x)=\hat{\Lambda}^{A}(x)T_{A}$, then, for some generic NC field $\hat{\Psi}$ that transforms in the fundamental representation of the gauge group (although the argument holds in any representation), we have
\begin{align}\label{NC_closure}
[\delta^{\star}_{1}\ds\delta^{\star}_{2}]\hat{\Psi}&=(\hat{\Lambda}_{1}\star\hat{\Lambda}_{2}
-\hat{\Lambda}_{2}\star\hat{\Lambda}_{1})\star\hat{\Psi} \nn\\
&= \frac{1}{2}\left([\hat{\L}^{A}_{1}\ds\hat{\L}_{2}^{B}]\{T_{A},T_{B}\}
+\{\hat{\L}^{A}_{1}\ds\hat{\L}_{2}^{B}\}[T_{A},T_{B}]\right)\star\hat{\Psi}=i\hat{\L}_{3}\star\hat{\Psi}=\delta^{\star}_{3}\hat{\Psi}\ .
\end{align}
The NC closure rule
\begin{equation}
[\delta^{\star}_{\hat{\L}_{1}}\ds\delta^{\star}_{\hat{\L}_{2}}]=\delta^{\star}_{-i[\hat{\L}_{1}\ds\hat{\L}_{2}]}
\end{equation} 
consistently generalizes its commutative counterpart.

However, (\ref{NC_closure}) implies that commutator of two NC gauge transformations does not generally close in the Lie algebra, because anti-commutator $\{T_{A},T_{B}\}$ does not in general belong to this algebra (except for $U(N)$ gauge group). To overcome this difficulty, we will apply the universal enveloping algebra (UEA) approach. Enveloping algebra is "large enough" to ensure that closure property of NC gauge transformations holds, provided that NC gauge parameter $\hat{\Lambda}$ is UEA-valued.

\noindent
NC covariant derivative (for a generic gauge group $\mathcal{G}$) in the fundamental representation is defined by
\begin{equation}
D_{\m}\hat{\Psi}=\partial_{\m}\hat{\Psi}-i\hat{V}_{\m}\star\hat{\Psi}\ ,
\end{equation}
where $\hat{V}_{\m}$ stands for the corresponding NC gauge field, and it transforms as
\begin{equation}
\delta^{\star}_{\L}D_{\m}\hat{\Psi}=i\hat{\L}\star D_{\m}\hat{\Psi}\ ,
\end{equation}
implying 
\begin{equation}
\delta^{\star}_{\L}\hat{V}_{\m}=\partial_{\m}\hat{\L}+i[\hat{\L}\ds\hat{V}_{\m}]\ .
\end{equation}
Therefore, NC gauge field must also be UEA-valued and it can be represented in its basis. But, UEA has an infinite basis, and it seems that by invoking it we actually introduced an infinite number of new degrees of freedom (new fields) in the NC theory, rendering it unrealistic. This problem is resolved by the Seiberg-Witten (SW) map \cite{SWMap1,SWMap2, SWMap3, SWMap4}. Essentially, we assume that classical gauge
transformations induce the corresponding NC gauge transformations,
\begin{equation}
\delta^\star_\L \hat {V}_{\m}=\hat{V}_{\m}(V_{\m}+\delta_{\epsilon}V_{\m})-\hat{V}_{\m}(V_{\m})\ .
\end{equation}
This allows us to represent every NC fields as a perturbation series in powers of the deformation parameter $\theta^{\m\n}$ with expansion coefficients built out of commutative fields, e.g. $\hat{\L}_{\epsilon}=\epsilon+\hat{\L}^{(1)}+\hat{\L}^{(2)}+...$. At zeroth order, NC fields reduce to their undeformed counterparts. For example, NC gauge parameter and potential can be represented as
\begin{align}
\hat{\L}_{\epsilon}&=\epsilon
-\frac14\theta^{\r\s}\{V_\r,\pa_\s\epsilon\}
+{\cal O}(\theta^2)\ , \\
\hat{V}_\m&= V_\mu
-\frac{1}{4}\theta^{\r\s}\{V_\r, \partial_\s V_\mu +
F_{\s\m}\}
+ {\cal O}(\theta^2)\ .
\end{align}

After these general considerations, we return to the NC action (\ref{NC42}). Field strength $\hat{\mathbb{F}}_{\m\n}$ appearing in (\ref{NC42}) is defined in terms of $OSp(4,2)_{\star}$ gauge potential 
$\hat{\Omega}_{\m}$ as
\be \hat{\mathbb{F}}_{\m\n}=\pa_\m\hat{\Omega}_\n-\pa_\n\hat{\Omega}_\m
-i[\hat{\Omega}_\m\ds\hat{\Omega}_\n] \ . 
\label{nckrivina}
\ee 
It transforms in the adjoint representation of $OSp(4,2)_{\star}$ supergroup as well as the NC auxiliary field $\hat{\Phi}$, 
\begin{align}
\delta^\star_\epsilon \hat {\mathbb{F}}_{\m\n}=i[\hat{\L}_{\epsilon} 
\ds \hat
{\mathbb{F}}_{\m\n}]\ ,\;\;\; \delta^\star_\epsilon \hat {\Phi}=i[\hat{\L}_{\epsilon} 
\ds \hat
{\Phi}]\ .
\end{align}
At this point it would be tempting to proceed by directly imposing the physical gauge. However, this operation would not yield an action with an appropriate symmetry because gauge fixing does not commute with NC deformation. A bypass is provided by the SW map. Representing NC fields in terms of commutative ones, as prescribed by the SW map, we obtain a perturbative expansion of $OSp(4\vert 2)_{*}$ gauge-invariant NC action (\ref{NC42}) in powers of the deformation parameter $\theta^{\m\n}$. By construction, SW map ensures invariance of the expansion under ordinary $OSp(4\vert 2)$ gauge transformations, order-by-order.
%%%%%%%%%%%%%%%%%%%%%%%%%%%%%%%%%%%%%%%%%%%%%

Now we present some relevant steps in the expansion procedure of the NC action (\ref{NC42}). Our goal is to calculate and analyze linear NC correction to the classical action (\ref{KomDejstvo(42)}). According to the SW map, the first order NC corrections of the auxiliary field $\Phi$ and the $OSp(4\vert 2)$ field strength $\mathbb{F}_{\m\n}$ are given by 
%%%%%%%%%%%%%%%%%%%%%%%%%%%
\be
\hat{\Phi}^{(1)}=-\frac{1}{4}\theta^{\r\s}
\{\Omega_\r, (\pa_\s+\widehat{D}_\s)\Phi\}\ ,
\ee
\begin{equation}
\hat{\mathbb{F}}_{\m\n}^{(1)}=-\frac{1}{4}\theta^{\r
\s}\{\O_{\r},(\partial_{\s}+\widehat{D}_{\s})\mathbb{F}_{\m
\n}\}+\frac{1}{2}\theta^{\r\s}\{\mathbb{F}_{\r\m},\mathbb{F}_{\s\n}\}\ , 
\end{equation}
where $\widehat{D}_{\m}$ stands for the $OSp(4|2)$ covariant derivative (associated to $\O_{\m}$).

Generally, for a pair of NC fields
$\hat{A}$ and $\hat{B}$, 
the linear NC correction to their product is
\begin{equation} 
\left(\hat{A}\star\hat{B}\right)^{(1)}
=\hat{A}^{(1)}B+A\hat{B}^{(1)}
+\frac{i}{2}\theta^{\r\s}\partial_\r A \partial_\s B \ . \label{rule}
\end{equation}
In particular, if both fields transform in the adjoint representation of $OSp(4|2)_{\star}$, we have 
\begin{align}
 \left(\hat{A}\star\hat{B}\right)^{(1)}=
 &-\frac{1}{4}\theta^{\r\s}\{\Omega_\r,(\partial_\s+\widehat{D}_\s)AB\}
 +\frac{i}{2}\theta^{\r\s} \widehat{D}_\r A \widehat{D}_\s B \nn \\
 &\hspace{0.4cm}+cov(\hat{A}^{(1)})B+Acov(\hat{B}^{(1)}) \ , \label{rule1}
\end{align}    
where $cov(\hat{A}^{(1)})$ is the covariant part of $A's$ first order NC correction, and $cov(\hat{B}^{(1)})$, 
the covariant part of $B's$ first order NC correction. Successive application of this rule gives us 
the first order NC correction to the classical action (\ref{KomDejstvo(42)}):
%%%%%%%%%%%%%%%%%%%%%%%%%%%%
\begin{align}
S^{(1)}_{42}=\frac{il\theta^{\l\t}}{32\pi G_{N}}\Str\int\diff^{4}x\;\varepsilon^{\m\n\r\s}\Bigg(&-\frac{1}{4}\{\mathbb{F}_{\l\t},\mathbb{F}_{\m\n}\mathbb{F}_{\r\s}\}\Phi
+\frac{i}{2}\widehat{D}_{\l}\mathbb{F}_{\m\n}\widehat{D}_{\t}
\mathbb{F}_{\r\s}\Phi \nn\\
&+\frac{1}{2}\{\mathbb{F}_{\l\m},\mathbb{F}_{\t\n}\}\mathbb{F}_{\r\s}\Phi
+\frac{1}{2}\mathbb{F}_{\m\n}
\{\mathbb{F}_{\l\r},\mathbb{F}_{\t\s}\}\Phi \nn\\
-\frac{1}{2l^{2}}\Big(&-\frac{1}{4}\{\SF_{\l\t},\SF_{\m\n}\Phi^{2}\}\SF_{\r\s}\Phi
+\frac{i}{2}\widehat{D}_{\l}\SF_{\m\n}\widehat{D}_{\t}\Phi^{2}\SF_{\r\s}\Phi \nn \\
&+\frac{1}{2}\{\SF_{\l\m},\SF_{\t\n}\}\Phi^{2}\SF_{\r\s}\Phi 
+\frac{i}{4}\SF_{\m\n}[\widehat{D}_{\l}\Phi,\widehat{D}_{\t}\Phi]\SF_{\r\s}\Phi \nn \\ 
&+\frac{i}{2}\SF_{\m\n}\Phi^{2}\widehat{D}_{\l}\SF_{\r\s}\widehat{D}_{\t}\Phi
+\frac{1}{2}\SF_{\m\n}\Phi^{2}\{\SF_{\l\r},\SF_{\t\s}\}\Phi\Big)
\Bigg)\ .
\end{align}
This linear NC correction is real and invariant under $OSp(4,2)$ gauge transformations. However, a careful examination shows that after the gauge fixing it vanishes completely,
\begin{equation}
S^{(1)}_{42}|_{g.f.}=0\ .
\end{equation}
But we still have the additional NC action $S^{\star}_{A}$ invariant under purely bosonic NC-deformed $SO(2,3)_{\star}\times U(1)_{\star}$ gauge transformations. The only additional SW expansion we need is that of $\hat{f}$, namely
\begin{equation}
\hat{f}=f-\frac{1}{4}\theta^{\r\s}
\{\Omega_\r, (\pa_\s+D_\s)f\}+\mathcal{O}(\theta^{2})\ .
\end{equation}
The first order NC correction to (\ref{NCSA}) before gauge fixing is given by
\begin{align}
S^{(1)}_{A}=&S^{(1)}_{Af}+S^{(1)}_{Aff} \nn\\  
=&-\frac{\theta^{\l\t}}{64l}\;\tr\int{\rm d}^{4}x\;\varepsilon^{\m\n\r\s}\bigg(-if D_\l \widetilde{f}_{\m\n}D_\t(D_\r\phi D_\s\phi \phi)
+\frac{1}{2}\{\widetilde{f}_{\l\t}, f\}\widetilde{f}_{\m\n}D_\r\phi D_\s\phi \phi  \nn\\
& - f\{\widetilde{f}_{\l\m},\widetilde{f}_{\t\n}\}D_\r\phi D_\s\phi \phi 
-if\widetilde{f}_{\m\n}D_\l(D_\r\phi D_\s\phi)D_\t\phi \nn\\
&-if\widetilde{f}_{\m\n}(D_\l D_\r\phi)(D_\t D_\s\phi)\phi \nn - f\widetilde{f}_{\m\n}\{\{\widetilde{f}_{\l\r},D_\t\phi\}, D_\s\phi\}\phi \nn\\[0.2cm]
& +\frac{i}{3!}\bigg(\frac{1}{2}\{\widetilde{f}_{\l\t}, f^{2}\} D_\m\phi D_\n\phi D_\r\phi D_\s \phi \phi - f^{2}\{[\{\widetilde{f}_{\l\m}, D_\t\phi\}, D_\n\phi], D_\r\phi D_\s\phi\}\phi \nn\\
&-if^{2}\Big(D_\l(D_\m\phi D_\n\phi D_\r\phi D_\s\phi)D_\t\phi
+D_\l(D_\m\phi D_\n\phi D_\r\phi)(D_\t D_\s\phi)\phi \nn\\
&+D_\l(D_\m \phi D_\n \phi)(D_\t D_\r\phi))D_\s\phi\phi
+(D_\l D_\m\phi)(D_\t D_\n\phi)D_\r\phi D_\s\phi\phi\Big)\bigg)\bigg)+ c.c. \label{NCSASW}
\end{align}
where we can distinguish the linear $f$-part and the quadratic $f^{2}$-part, and all terms are manifestly $SO(2,3)\times U(1)$ invariant by the virtue of SW map. 

After calculating traces and evaluating the gauge-fixed action $S^{(1)}_{A}|_{\text{g.f.}}$ on the EoMs of the components of the auxiliary field $f$ (as it turns out, to obtain the first order NC correction, we only need to insert \emph{zeroth order} (classical) EoMs (\ref{EoMs}) in the gauge-fixed \emph{first order} NC action $S^{(1)}_{A}|_{\text{g.f.}}$), we obtain 
\begin{equation}
S^{(1)}_{A,EoM}|_{\text{g.f.}}=\sum\limits_{j=1}^6 S^{(1)}_{A,EoMf.j}|_{\text{g.f.}}+S^{(1)}_{A,EoMff}|_{\text{g.f.}}\ ,\label{NCSA1EOM} 
\end{equation}
with the individual terms:
\begin{align}
&S^{(1)}_{A,EoMf.1}|_{\text{g.f.}}=-\frac{\theta^{\l\t}}{64\kappa}\int{\rm d}^{4}x\;e\;
\Bigg\{\widetilde{\mathcal{F}}^{\m\n}R_{\m\n ab}\left(R_{\l\t}^{\;\;\;\;ab}-\frac{2}{l^2}e_{\l}^{a}e_{\t}^{b}\right) \nn\\
&+\widetilde{\mathcal{F}}^{\r\s}e_{\r}^{a}e_{\s}^{b}\left(R_{\m\n ab}R_{\l\t}^{\;\;\;\;cd}e_{c}^{\m}e_{d}^{\n}-\frac{2}{l^2}R_{\l\t ab}\right)
+4\widetilde{\mathcal{F}}^{\r\m}e_{\r}^{c}\left(R_{\m\n ac}R_{\l\t}^{\;\;\;\;ab}e_{b}^{\n}+\frac{2}{l^2}R_{\m\l ac}e_{\t}^{a}\right) \nn\\
&+\widetilde{\mathcal{F}}_{\r\s}e^{\r}_{a}e^{\s}_{b}R_{\l\t}^{\;\;\;\;ab}
\left(R_{\m\n}^{\;\;\;\;mn}e_{m}^{\m}e_{n}^{\n}-\frac{12}{l^2}\right)
+\frac{2}{l^2}\widetilde{\mathcal{F}}^{\m\n}T_{\l\t}^{\;\;\;\;a}\big( T_{\m\n a}-2T_{\r\n m}e_{a}^{\r}e_{\m}^{m}\big)\nn\\
&-\frac{2}{l^2}\widetilde{\mathcal{F}}_{\l\t}\left(R_{\m\n}^{\;\;\;\;mn}e_{m}^{\m}e_{n}^{\n}-\frac{12}{l^2}-4\frac{l^{2}}{\kappa^{2}}\widetilde{\mathcal{F}}^{\m\n}\widetilde{\mathcal{F}}_{\m\n}\right)\Bigg\}\ , \label{1112}
\end{align}
\begin{align}
&S^{(1)}_{A,EoMf.2}|_{\text{g.f.}}=-\frac{\theta^{\l\t}}{8\kappa}\int {\rm d}^{4}x\;e\;\Bigg\{\nn\\
&+(D^{L}_\l R_{\m\n}^{\;\;\;\;mc})(D^{L}_\t e_{\r}^{r})e_{c}^{\s}\left( e_{m}^{\n}(\widetilde{\mathcal{F}}_{\s}^{\;\;\m}e_{r}^{\r} - \widetilde{\mathcal{F}}_{\s}^{\;\;\r}e_{r}^{\m}) + \widetilde{\mathcal{F}}_{\s}^{\;\;\n}e_{r}^{\m}e_{m}^{\r}\right)
\nn\\
&-\frac{1}{l^2}\widetilde{\mathcal{F}}_{\r}^{\;\;\m}e_{c}^{\r}\big(D^{L}_\l T_{\t\m}^{\;\;\;\;c}-e_{\l b}R_{\t\m}^{\;\;\;\;bc}\big)-\frac{4}{l^2}\widetilde{\mathcal{F}}_{\n}^{\;\;\m}(D^{L}_\l e_{\r}^{r})(D^{L}_\t e_\m^m)e_{m}^{\n}e_{r}^{\r} \nn\\
&-\frac{1}{l^2}(D^{L}_\l e_{\r}^{r})e_{r}^{\r}\left( e_{c}^{\n}\widetilde{\mathcal{F}}_{\t}^{\;\;\m}T_{\m\n}^{\;\;\;\;c}-e_{c}^{\n}\widetilde{\mathcal{F}}_{\n}^{\;\;\m}T_{\m\t}^{\;\;\;\;c}\right)
+\frac{1}{2l^2}T_{\l\t}^{\;\;\;\;r}T_{\m\n}^{\;\;\;\;c}\widetilde{\mathcal{F}}_{\s}^{\;\;\n}e_{c}^{\s}e_{r}^{\m} \nn\\
& +\frac{1}{l^2}(D^{L}_\l e_{\r}^{r})e_{r}^{\n} \Big( T_{\m\n}^{\;\;\;m}\big( \widetilde{\mathcal{F}}_{\t}^{\;\;\m}e_{m}^{\r} - \widetilde{\mathcal{F}}_{\t}^{\;\;\r}e_{m}^{\m}\big) + T_{\t\n}^{\;\;\;m}\widetilde{\mathcal{F}}_{\m}^{\;\;\r}e_{m}^{\m} \Big)\nn\\
& +\frac{2}{l^2} \widetilde{\mathcal{F}}_{\s}^{\;\;\r}e_{c}^{\s}(D^{L}_\l e_{\r}^{r})(D^{L}_\t e_\n^c)e_{r}^{\n} +\frac{1}{l^4}\widetilde{\mathcal{F}}_{\l\t}\Bigg\}\ , \label{13}
\end{align}
\begin{align}
&S^{(1)}_{A,EoMf.3}|_{\text{g.f.}}=\frac{\theta^{\l\t}}{32\kappa}\int {\rm d}^{4}x\;e\; \Bigg\{-\widetilde{\mathcal{F}}^{\m\n}R_{\l\n am}\left( R_{\t\m}^{\;\;\;\;am} 
-\frac{4}{l^2}e_\t^a e_\m^m\right)\nn\\
&+\widetilde{\mathcal{F}}_{\r\s}R_{\l\m}^{\;\;\;\;am}R_{\t\n}^{\;\;\;\;bn} \big( e_{a}^{\m}e_{m}^{\n}e_{b}^{\r}e_{n}^{\s}+e_{a}^{\r}e_{m}^{\s}e_{b}^{\m}e_{n}^{\n}+2e_{n}^{\r}e_{m}^{\s}(e_{a}^{\m}e_{b}^{\n}-e_{a}^{\n}e_{b}^{\m})\big)\nn\\
&-\frac{2}{l^2}e_{n}^{\r}e_{b}^{\s}\left(2\widetilde{\mathcal{F}}_{\l\r}R_{\t\s}^{\;\;\;\;bn}+\widetilde{\mathcal{F}}_{\r\s}R_{\l\t}^{\;\;\;\;bn}\right) 
+\frac{2}{l^2}\widetilde{\mathcal{F}}^{\m\n}T_{\l\m}^{\;\;\;\;a}T_{\t\n a}+\frac{8}{\kappa^{2}}\widetilde{\mathcal{F}}^{\m\n}\widetilde{\mathcal{F}}_{\l\m}\widetilde{\mathcal{F}}_{\t\n}\Bigg\}\ , \label{14}
\end{align}
\begin{align}
&\left(S^{(1)}_{A,EoMf.4}|_{\text{g.f.}} + S^{(1)}_{A,EoMf.5}|_{\text{g.f.}}\right)=-\frac{\theta^{\l\t}}{32\kappa}\int {\rm d}^{4}x\;e\;\Bigg\{\nn\\
&+R_{\mu\nu}^{\;\;\;\;ab}(D^{L}_\l e_{\rho}^{m})(D^{L}_\t e_{\sigma m})\Big(\widetilde{\mathcal{F}}^{\m\n}e_{a}^\r e_{b}^\s 
+ \widetilde{\mathcal{F}}^{\r\s}e_{a}^\m e_{b}^\n -4\widetilde{\mathcal{F}}^{\m\r}e^{\n}_a e^{\s}_b\Big)\nn\\
&-\frac{1}{l^2}R_{\mu\nu}^{\;\;\;\;ab}\big(\widetilde{\mathcal{F}}^{\m\n}e_{\l a} e_{\t b} 
+ \widetilde{\mathcal{F}}_{\l\t}e_{a}^\m e_{b}^\n - 4\widetilde{\mathcal{F}}^\m_{\;\;\l}e^{\n}_a e_{\t b}\big) 
-\frac{1}{l^2}\widetilde{\mathcal{F}}^{\m\n}\eta_{mn}(D^{L}_{\l}e_{\m}^{m})(D^{L}_{\t}e_{\n}^{n})\nn\\
& +\frac{2}{l^2}T_{\mu\nu}^{\;\;\;a}(D^{L}_\l e_\rho^b)\Big( \widetilde{\mathcal{F}}^{\m\n}(2e^{\r}_a e_{\t b} - e^{\r}_b e_{\t a}) 
+ 2 \widetilde{\mathcal{F}}_\t^{\;\;\m}(e^{\r}_a e_b^\n  - e^{\r}_b e_a^\n)\nn\\
& +2\widetilde{\mathcal{F}}^{\r\m}(2e^{\n}_a e_{\t b}  - e^{\n}_b e_{\t a}) +2\widetilde{\mathcal{F}}^\r_{\;\;\t}e^{\m}_a e_b^\n\Big)  
-\frac{2}{l^2}\widetilde{\mathcal{F}}^{\r\m}T_{\mu\nu c}T_{\l\t}^{\;\;\;\;d}e_{\r}^c e_d^\n  \nn\\
& +\frac{4}{l^2}T_{\l\nu a}(D^{L}_\t e_\rho^d)e^{a}_\s\big( \widetilde{\mathcal{F}}^{\s\n}e_d^\r
-\widetilde{\mathcal{F}}^{\s\r}e_d^\n \big) 
-\frac{4}{l^2}R_{\l\nu\;\;a}^{\;\;\;\; c}e^{a}_\r\big( \widetilde{\mathcal{F}}^{\r\n}e_{\t c}
-\widetilde{\mathcal{F}}^\r_{\;\;\t}e_c^\n \big)
-\frac{6}{l^4}\widetilde{\mathcal{F}}_{\l\t} \Bigg\}\ ,  \label{1516}
\end{align}
\begin{align}
&S^{(1)}_{A,EoMf.6}|_{\text{g.f.}} = \frac{\theta^{\l\t}}{32\kappa}\int {\rm d}^{4}x\;e\;\Bigg\{\widetilde{\mathcal{F}}^{\r\s}e_{\r}^{a}e_{\s}^{b}e^{\m}_{m}e^{\n}_{n}\big( R_{\l\t}^{\;\;\;\;mn}R_{\mu\nu ab}-2R_{\l\m}^{\;\;\;\;mn}R_{\t\n ab} \big) \nn\\
&+\frac{8}{\kappa^{2}}\big(\widetilde{\mathcal{F}}_{\l\t}\widetilde{\mathcal{F}}^{2}-2\widetilde{\mathcal{F}}^{\m\n}\widetilde{\mathcal{F}}_{\l\m}\widetilde{\mathcal{F}}_{\t\n}\big) 
+\frac{2}{l^2}\widetilde{\mathcal{F}}^{\m\n}e_{\m}^{a}e_{\n}^{b}R_{\l\t ab}+\frac{4}{l^2}\widetilde{\mathcal{F}}_{\l\n}\e^{\n}_a e^{\r}_b R_{\t\rho}^{\;\;\;\;ab}  \nn\\
&-\frac{4}{l^2}\widetilde{\mathcal{F}}^{\m\n}e_{\n}^{a}e^{\r}_{b} \big( T_{\r\m a}T_{\l\t}^{\;\;\;\;b}
-T_{\m\l a}T_{\t\r}^{\;\;\;\;b} - T_{\l\r a}T_{\t\m}^{\;\;\;\;b}\big) - \frac{8}{l^4}\widetilde{\mathcal{F}}_{\l\t}\Bigg\}\ ,
\label{1718}
\end{align}       
And finally, the $f^{2}$-term,
\begin{equation}
S^{(1)}_{A,EoMff}|_{\text{g.f.}}=-\frac{\theta^{\l\t}}{16\kappa^{3}}\int {\rm d}^{4}x\;e\;\widetilde{\mathcal{F}}_{\l\t}\widetilde{\mathcal{F}}^{2} \ . 
\end{equation}
Action (\ref{NCSA1EOM}) represents the first order NC correction to $N=2$ $AdS_{4}$ SUGRA. It involves various new couplings between $U(1)$ gauge field, gravity and gravitini fields that appear due to space-time noncommutativity. As it stands, this action seems too complicated to be analyzed in its entirety. However, we can restrict ourselves to some particular domain of parameters and work with an approximated NC action. In particular, we will derive a low-energy approximation of (\ref{NCSA1EOM}), by taking into account terms at most quadratic in partial derivative. Therefore, we include only terms linear in curvature, 
and linear and quadratic in 
torsion. Additionally, we assume that spin connection $\omega_\mu^{\;\;ab}$ and the first order derivatives of vierbeins are of the same order. Note also that the torsion constraint $\widetilde{T}_{\m\n}^{\;\;\;\;a}=0$ (\ref{ST}) gives us $T_{\m\n}^{\;\;\;\;a}=-i\Psi_{\m}\g^{a}\Psi_{\n}$. These assumptions yield a very simple action, 
\begin{align}
S^{(1)}_{\text{low-energy}}&=-\frac{9\theta^{\m\n}}{16l^{4}\kappa}\int{\rm d}^{4}x\;e\;\widetilde{\mathcal{F}}_{\m\n}=-\frac{9\theta^{\m\n}}{16l^{4}\kappa}\int{\rm d}^{4}x\;e\;(\mathcal{F}_{\m\n}-\bar{\Psi}_{\m}i\s^{2}\Psi_{\n})\nn\\
&=\frac{9\theta^{\m\n}}{16l^{4}\kappa}\int{\rm d}^{4}x\;e\;(\bar{\psi}^{1}_{\m}\psi^{2}_{\n}-\bar{\psi}^{2}_{\m}\psi^{1}_{\n})+\text{surface term} \nn\\
&=-\frac{9}{8l^{4}\kappa}\int{\rm d}^{4}x\;e\;(\bar{\psi}^{+}_{\m}i\Theta^{\m\n}\psi^{+}_{\n})+\text{surface term}\ .
\end{align}
This mass-like term for charged gravitino $\psi^{+}_{\m}$, minimally coupled to gravity, appears due to space-time noncommutativity, and ``renormalizes'' the corresponding term (\ref{mass-like term}) in the classical SUGRA action (\ref{Final42classicalDirac}). If we again absorb $\kappa^{-1}$ in $\psi^{+}_{\m}$ to obtain the canonical dimensions, the mass-like parameter is $\sim l_{P}\Lambda_{NC}^{2}/l^{4}$, and it vanishes under WI contraction.

After WI contraction, the action (\ref{NCSA1EOM}) reduces to
\begin{align}\label{NC42SUGRA_linearWI}
&S_{A}|_{\text{g.f.}}\myeq-\frac{\theta^{\l\t}}{64\kappa}\int{\rm d}^{4}x\;e\;\Bigg\{\widetilde{\mathcal{F}}^{\m\n}R_{\m\n\r\s}R_{\l\t}^{\;\;\;\;\r\s}
-\tilde{\mathcal{F}}^{\m\n}R_{\r\s\m\n}R_{\l\t}^{\;\;\;\;\r\s}-4\widetilde{\mathcal{F}}^{\m\r}R_{\m\n\r\s}R_{\l\t}^{\;\;\;\; \n\s}\nn\\
&-2\widetilde{\mathcal{F}}^{\m\n}R_{\l\m}^{\;\;\;\;\r\s}R_{\t\n\r\s}+8\widetilde{\mathcal{F}}_{\r\s}R_{\l\m}^{\;\;\;\;\m\r}R_{\t\n}^{\;\;\;\;\n\s}+\widetilde{\mathcal{F}}_{\m\n}R_{\l\t}^{\;\;\;\;\m\n}R-\frac{4}{\kappa^{2}}\widetilde{\mathcal{F}}_{\l\t}\widetilde{\mathcal{F}}^{2}+\frac{16}{\kappa^{2}}\widetilde{\mathcal{F}}^{\m\n}\widetilde{\mathcal{F}}_{\l\m}\widetilde{\mathcal{F}}_{\t\n}\nn\\
&+8(D^{L}_\l R_{\m\n}^{\;\;\;mc})(D^{L}_\t e_{\r}^{r})e_{c}^{\s} \Big(\widetilde{\mathcal{F}}_{\s}^{\;\;\m}e_{m}^{\n}e_{r}^{\r}-\widetilde{\mathcal{F}}_{\s}^{\;\;\r}e_{r}^{\m}e_{m}^{\n}+\widetilde{\mathcal{F}}_{\s}^{\;\;\n}e_{r}^{\m}e_{m}^{\r}\Big)\nn\\
&+2R_{\mu\nu}^{\;\;\;\;ab}\eta_{rs}(D^{L}_\l e_{\rho}^{r})(D^{L}_\t e_{\sigma}^{s})\Big(\widetilde{\mathcal{F}}^{\m\n}e_{a}^\r e_{b}^\s
+\widetilde{\mathcal{F}}^{\r\s}e_{a}^\m e_{b}^\n
-4\widetilde{\mathcal{F}}^{\m\r}e^{\n}_a e^{\s}_b\Big)\Bigg\}\ .
\end{align}
At this point we are confronted with an interesting question. The fact that $N=2$ $AdS_{4}$ superalgebra contracts to $N=2$ Poinacar\'{e} superalgebra when $l\rightarrow\infty$ is consistently reflected on the level of \emph{classical} (undeformed) action (\ref{Final42classicalMajorana}); classical $N=2$ $AdS_{4}$ SUGRA reduces to classical $N=2$ Poincar\'{e} SUGRA under WI contraction. However, it is not a priori clear whether this relation holds \emph{after NC deformation}, that is, whether NC deformation and WI contraction actually commute. For that matter, one would have to explicitly compute the NC correction to classical $N=2$ Poincar\'{e} SUGRA and compare it to the action (\ref{NC42SUGRA_linearWI}).

%%%%%%%%%%%%%%%%%%%%%%%%%%%%%%%%%%%%%%%%%%%%%
\section{Discussion and Outlook}

Let us emphasize the main points of this paper and propose some further paths of investigation. At this stage, our primary goal was to obtain explicit NC correction to $N=2$ AdS SUGRA in $D=4$. We stared from a classical (undeformed) action (\ref{KomDejstvo(42)}) of the Yang-Mills type (already advocated in the literature), invariant under orthosymplectic $OSp(4\vert 2)$ gauge transformations. However, this action alone is not enough to obtain $N=2$ $AdS_{4}$ SUGRA after fixing the gauge (for which we use a constrained auxiliary field). In particular, one has to add a supplementary action (\ref{SA_SUGRA}) endowed with $SO(2,3)\times U(1)$ gauge symmetry (bosonic sector of $OSp(4|2)$) that provides the missing terms in the classical SUGRA action (e.g. the kinetic term for $U(1)$ gauge field). Therefore, we have the following schema:
\begin{equation*}
\begin{matrix}  
(OSp(4|2)\; \text{invariant action})&+&(SO(2,3)\times U(1)\; \text{invariant action}) & \cr  \arrow{d}{\text{\large g.f.}} &           & \arrow{d}\;{\text{\large g.f.}} \cr (SO(1,3)\times U(1)\; \text{invariant action})      & + & (SO(1,3)\times U(1)\; \text{invariant action}) \cr
\end{matrix}
\end{equation*}
\begin{equation*}
\underbrace{\;\;\;\;\;\;\;\;\;\;\;\;\;\;\;\;\;\;\;\;\;\;\;\;\;\;\;\;\;\;\;\;\;\;\;\;\;\;\;\;\;\;\;\;\;\;\;\;\;\;\;\;\;\;\;\;\;\;\;\;\;\;\;\;\;\;\;\;\;\;\;\;\;\;\;\;\;\;\;\;\;\;\;\;\;\;\;\;\;\;\;\;\;\;\;\;\;\;\;\;\;\;\;\;\;\;\;\;\;\;\;\;\;\;\;\;\;\;\;}_\text{\large N=2 AdS SUGRA in D=4}
\end{equation*}

This situation seems curious considering that a similar $OSp(4|1)$ gauge-invariant action (\ref{KomDejstvo(41)}) in the same gauge reduces  to the complete $N=1$ $AdS_{4}$ SUGRA action. We may conclude that extended $N>1$ $AdS_{4}$ SUGRA cannot be obtained simply by gauging the corresponding orthosymplectic group $OSp(4|N)$ and subsequently fixing the gauge. For $N>1$ we would have to include an additional term similar to (\ref{SA_SUGRA}) that involves non-Abelian Yang-Mills gauge field. 
  
NC deformation is performed following the Seiberg-Witten approach to NC gauge field theory that involves universal enveloping algebra-valued gauge field and perturbative expansion of the NC-deformed action in powers of the deformation parameter $\theta^{\m\n}$. The expanded action possesses gauge symmetry of the corresponding classical action, order-by-order, and we focus only on the linear NC correction that remains after the gauge fixing. For the $OSp(4|2)$ gauge-invariant part, linear NC correction vanishes.
The reason why this result strikes us as curious is related to some previously establish facts about canonical NC deformation of the similar models. Namely, canonical deformation of pure gravity, regarded as a gauge theory of $SO(2,3)$ group, leads to quadratic NC correction \cite{MDVR-14, Us-16, PLM-13}. However, after including charged matter (Dirac spinors) coupled to $U(1)$ gauge field, linear NC correction appears \cite{VG, U(1), Paolo-fer}. Since we can take a pair of Majorana vector-spinors of $OSp(4|2)$ SUGRA and form a pair of $U(1)$-charged Dirac vector-spinors, related to each other by $C$-conjugation, we expected to obtain a non-vanishing first order NC correction from the $OSp(4|2)$ action (\ref{NC42}), as well.      
 
However, the supplementary bosonic action provides a non-trivial linear NC correction that is calculated explicitly (\ref{NCSA1EOM}). It involves various new interaction terms that are present due to space-time noncommutativity. The full action is difficult to analyze, but we can restrict ourselves to the low-energy sector of the theory by taking into account only terms that are at most quadratic in partial derivatives. This leaves us with a single mass-like term for charged gravitino minimally coupled to gravity.

WI contraction eliminates many of these new interaction terms, but not all of them (\ref{NC42SUGRA_linearWI}). The ones remaining may help us understand the relation between the canonical NC deformation and WI contraction, at least in this particular case. $N=2$ AdS superalgebra reduces to 
$N=2$ Poincar\'{e} superalgebra under WI contraction and the same holds for their classical actions. Therefore, it may be the case that the same relation pertains even after canonical NC deformation.    
To confirm this assumption directly, we have to calculate linear NC correction to $N=2$ Poincar\'{e} SUGRA, and make the comparison.  

Let us just mention that there are additional two terms with $OSp(4|2)$ gauge symmetry that we could include. We denote them by $S'$ and $S''$ and they are given by
\begin{align}
S'&=\frac{a'}{128 \pi G_{N}l}\Str\int \diff^{4}x\;\varepsilon^{\m\n\r\s}\mathbb{F}_{\m\n}\widehat{D}_{\r}\Phi\widehat{D}_{\s}\Phi\Phi+c.c. \ , \nn \\
S''&=-\frac{ia''}{128 \pi G_{N}l^{3}}\Str\int \diff^{4}x\;\varepsilon^{\m\n\r\s}\widehat{D}_{\m}\Phi\widehat{D}_{\n}\Phi\widehat{D}_{\r}\Phi\widehat{D}_{\s}\Phi\Phi \ , \nn
\end{align}
with free dimensionless parameters $a'$, $a''$, and $OSp(4|2)$ covariant derivative $\widehat{D}_{\m}$. Their $SO(2,3)$ gauge-invariant counterparts were analyzed in \cite{Us-16}. After gauge fixing, they modify the coefficients in the classical action but do not introduce new terms. In particular, they give us a freedom to eliminate the cosmological constant in the classical action. NC deformation of $S'$ and $S''$ will certainly change our final result, but their importance is not yet clear. Analysis of these additional NC corrections remains to be done.

\vskip1cm \noindent 
{\bf Acknowledgement}    
We thank Maja Buri\'{c} for fruitful discussions and Aleksandra Dimi\'{c} for her valuable assistance. The work is supported by project ON171031 of the Serbian Ministry of Education and Science and partially supported by the Action MP1405 QSPACE from the Europe
an Cooperation in Science
and Technology (COST).
%%%%%%%%%%%%%%%%%%%%%%%%%%%%%%%%%%%%%%%%%%%%%%%%%%%%%%%%%%%%%%%%%%%%%%
\newpage
\initiate
\section{Appendix A} 
%%%%%%%%%%%%%%%%%%%%%%%%%%%%%%%%5
Here we present an explicit $6\times 6$ matrix representation of $OSp(4\vert 2)$ generators $\{\hat{M}_{AB},\hat{Q}_{\a}^{I},\hat{T}\}$. The AdS generators are the same as for $OSp(4|1)$, but for $OSp(4\vert 2)$ we have an additional set of fermionic generators (comprising another Majorana spinor) and additional bosonic generator $\hat{T}$ of $SO(2)\sim U(1)$.

\textbf{Bosonic generators of} $OSp(4\vert 2)$:

\begin{equation}
\sbox1{$\begin{matrix}0&0&0&0 \end{matrix}$}
\sbox2{$\begin{matrix}0\\0\\0\\0 \end{matrix}$}
\hat{M}_{AB}=\left(
\begin{array}{c|cc}
\makebox[\wd0]{\large $M_{AB}$} & \usebox{2} & \usebox{2} \\
\hline
  \usebox{1} & 0 & 0 \\
  \usebox{1} & 0 & 0
\end{array}
\right) \;\;\;\; \hat{T}=\left(
\begin{array}{c|cc}
\makebox[\wd0]{\large $0_{4\times 4}$} & \usebox{2} & \usebox{2} \\
\hline
  \usebox{1} & 0 & i \\
  \usebox{1} & -i & 0
\end{array}
\right) \label{OSp42B}
\end{equation}  

The imaginary unit in $\hat{T}$ is introduced for convenience.\\[0.2cm]

\textbf{Fermionic generators of} $OSp(4\vert 2)$:

\begin{equation}
\sbox1{$\begin{matrix}1&0&0&0 \end{matrix}$}
\sbox2{$\begin{matrix}0\\1\\0\\0 \end{matrix}$}
\sbox3{$\begin{matrix}0&1&0&0 \end{matrix}$}
\sbox4{$\begin{matrix}-1\\0\\0\\0 \end{matrix}$}
\sbox5{$\begin{matrix}0\\0\\0\\0 \end{matrix}$}
\sbox6{$\begin{matrix}0&0&0&0 \end{matrix}$}
(\hat{Q}^{1})_{1}=\left(
\begin{array}{c|cc}
\makebox[\wd0]{\large $0_{4\times 4}$} & \usebox{2} & \usebox{5} \\
\hline
\usebox{1} & 0 & 0 \\
\usebox{6} & 0 & 0
\end{array}
\right)\ , \;\;\;\;
(\hat{Q}^{1})_{2}=\left(
\begin{array}{c|cc}
\makebox[\wd0]{\large $0_{4\times 4}$} & \usebox{4} & \usebox{5} \\
\hline
\usebox{3} & 0 & 0 \\
\usebox{6} & 0 & 0
\end{array}
\right) \label{OSp42F1}
\end{equation}

\begin{equation}
\sbox1{$\begin{matrix}0&0&1&0 \end{matrix}$}
\sbox2{$\begin{matrix}0\\0\\0\\-1 \end{matrix}$}
\sbox3{$\begin{matrix}0&0&0&1 \end{matrix}$}
\sbox4{$\begin{matrix}0\\0\\1\\0 \end{matrix}$}
\sbox5{$\begin{matrix}0\\0\\0\\0 \end{matrix}$}
\sbox6{$\begin{matrix}0&0&0&0 \end{matrix}$}
(\hat{Q}^{1})_{3}=\left(
\begin{array}{c|cc}
\makebox[\wd0]{\large $0_{4\times 4}$} & \usebox{2} & \usebox{5} \\
\hline
 \usebox{1} & 0 & 0 \\
 \usebox{6} & 0 & 0
\end{array}
\right) \;\;\;\;
(\hat{Q}^{1})_{4}=\left(
\begin{array}{c|cc}
\makebox[\wd0]{\large $0_{4\times 4}$} & \usebox{4} & \usebox{5} \\
\hline
\usebox{3} & 0 & 0 \\
\usebox{6} & 0 & 0
\end{array}
\right) \label{OSp42F2}
\end{equation}

The second set of fermionic generators $(\hat{Q}^{2})_{\a}$ $(\a=1,2,3,4)$ is obtained from the first one just by interchanging $5^{th}$ and $6^{th}$ column, and $5^{th}$ and $6^{th}$ row.   
One can readily check that supermatrices (\ref{OSp42B}), (\ref{OSp42F1}) and  (\ref{OSp42F2}), along with the second set of fermionic generators, satisfy the $\mathfrak{osp}(4\vert 2)$ superalgebra (\ref{OSp_algebra}).

\initiate
%%%%%%%%%%%%%%%%%%%%%%%%%%%%%%%%%
\section{Appendix B}
%%%%%%%%%%%%%%%%%%%%%%%%%%%%%%%%%
Some basic Fierz identities involving Majorana spinors $\psi$ and $\chi$:
\begin{align}
\bar{\psi}\chi&=\bar{\chi}\psi=(\bar{\psi}\chi)^{\dag} \nn\\
\bar{\psi}\gamma_{5}\chi&=\bar{\chi}\gamma_{5}\psi=-(\bar{\psi}\gamma_{5}\chi)^{\dag}\nn\\
\bar{\psi}\gamma_{a}\g_{5}\chi&=\bar{\chi}\gamma_{a}\gamma_{5}\psi=(\bar{\psi}\gamma_{a}\g_{5}\chi)^{\dag}\nn\\
\bar{\psi}\gamma_{a}\chi&=-\bar{\chi}\gamma_{a}\chi=-(\bar{\psi}\gamma_{a}\chi)^{\dag}\nn\\
\bar{\psi}\s_{ab}\chi&=-\bar{\chi}\s_{ab}\psi=-(\bar{\psi}\s_{ab}\chi)^{\dag}
\end{align}

Also, we frequently use the following important identity, valid in $4$D. For any pair of Majorana spinors, $\psi$ and $\chi$, we can expand $\psi\bar{\chi}$ in the Clifford algebra basis: 
\begin{equation}
-4\psi\bar{\chi}=(\bar{\chi}\psi)\mathbb{I}_{4}+(\bar{\chi}\g^{a}\psi)\g_{a}
+(\bar{\chi}\g_{5}\psi)\g_{5}+(\bar{\chi}\g^{a}\g_{5}\psi)\g_{5}\g_{a}
+\frac{1}{2}(\bar{\chi}\s^{ab}\psi)\s_{ab}
\end{equation}

\noindent\textbf{Some AdS algebra relations}\footnote{$\epsilon^{01235}=+1,\ \epsilon^{0123}=+1$}:
\bea
&&[M_{AB},M_{CD}]=i(\eta_{AD}M_{BC}+\eta_{BC}M_{AD}-\eta_{AC}M_{BD}-\eta_{BD}M_{AC})\nn\\
&&\{M_{AB},M_{CD}\}=\frac{i}{2}\epsilon_{ABCDE}\Gamma^{E}+\frac12(\eta_{AC}\eta_
{BD}-\eta_{AD}\eta_{BC})\nn\\
&&\{M_{AB},\Gamma_C\}=i\epsilon_{ABCDE}M^{DE}\nn\\
&&{[}M_{AB},\Gamma_C{]}=i(\eta_{BC}\Gamma_A-\eta_{AC}\Gamma_B)\nn\\
&&\Gamma_A^\dagger=-\gamma_0\Gamma_A\gamma_0\ ,\;\;\;\;M_{AB}^\dagger=\gamma_0M_{AB}\gamma_0 
\eea
Some useful identities involving $\g$-matrices and $\s$-matrices:
\bea
&&\g_a\g_b=\eta_{ab}-i\s_{ab}\nn\\
&&\s_{ab}\g_c=i\eta_{bc}\g_a-i\eta_{ac}\g_b+\varepsilon_{abcd}\g_5\g^d \nn\\
&&\g_c\s_{ab}=i\eta_{ac}\g_b-i\eta_{bc}\g_a+\varepsilon_{abcd}\g_5\g^d\nn\\
&&\sigma_{ab}\gamma_5=\tfrac{i}{2}\varepsilon_{abcd} \sigma^{cd}\nn\\
&&\sigma_{ab}\sigma_{cd}=\eta_{ac}\eta_{bd}-\eta_{ad}\eta_{bc}+i\varepsilon_{abcd}\gamma_5
  +i(\eta_{ad}\sigma_{bc}+\eta_{bc}\sigma_{ad}-\eta_{ac}\sigma_{bd}-\eta_{bd}\sigma_{ac})
\eea

\noindent\textbf{Identities with traces}:
\bea
&&\tr (\Gamma_A\Gamma_B)=4\eta_{AB}\nn\\
&&\tr (\Gamma_A)=\tr (\Gamma_A\Gamma_B\Gamma_C)=0\nn\\
&&\tr
(\Gamma_A\Gamma_B\Gamma_C\Gamma_D)=4(\eta_{AB}\eta_{CD}-\eta_{AC}\eta_{BD}+\eta_
{AD}\eta_{CB})\nn\\
&&\tr (\Gamma_A\Gamma_B\Gamma_C\Gamma_D\Gamma_E)=-4i\epsilon_{ABCDE}\nn\\
&&\tr (M_{AB}M_{CD}\Gamma_E)=i\epsilon_{ABCDE}\nn\\
&&\tr (M_{AB}M_{CD})=-\eta_{AD}\eta_{CB}+\eta_{AC}\eta_{BD} \nn \\ 
&&Tr(M_{AB}\Gamma_{E}\Gamma_{F}\Gamma_{G})=2\varepsilon_{ABEFG} \nn \\
&&Tr(M_{AB}M_{CD}\Gamma_{E}\Gamma_{F}\Gamma_{G})=i\varepsilon_{ABCDE}\eta_{FG}-i\varepsilon_{ABCDF}\eta_{EG}+i\varepsilon_{ABCDG}\eta_{EF} \nn\\
&&\;\;\;\;\;\;\;\;\;\;\;\;\;\;+i\varepsilon_{BCEFG}\eta_{AD}+i\varepsilon_{ADEFG}\eta_{BC}-i\varepsilon_{BDEFG}\eta_{AC}-i\varepsilon_{ACEFG}\eta_{BD} \label{AdStraces}
\eea

\newpage
%%%%%%%%%%%%%%%%%%%%%%%%%%%%%%%%%%%%%

\end{document}